\documentclass[a4paper, 11pt, oneside]{article} 

\usepackage{geometry}
\usepackage{booktabs} 

\usepackage{mathptmx} 

\usepackage{fancyhdr}
\usepackage[normalem]{ulem}
\usepackage{microtype}
\usepackage{graphicx}
\usepackage{colortbl}
\definecolor{mygray}{gray}{.88}
\usepackage{multirow}
\usepackage{amsmath}
\usepackage{bm}
\usepackage{epsfig}
\usepackage{authblk}

\usepackage[hyphens]{url}
\usepackage{hyperref}
\usepackage{color}
\usepackage{soul}
\usepackage{bbding}
\definecolor{mygray}{gray}{.88}

\usepackage[english]{babel}
\usepackage{blindtext}
\usepackage{multirow}
\usepackage{pifont}
\usepackage{subfig}
\usepackage{hyperref}
\usepackage{threeparttable}
\usepackage{makecell}
\usepackage{tabularx}
\newcolumntype{L}[1]{>{\raggedright\arraybackslash}p{#1}}
\newcolumntype{C}[1]{>{\centering\arraybackslash}p{#1}}
\newcolumntype{R}[1]{>{\raggedleft\arraybackslash}p{#1}}

\begin{document} 

\begin{titlepage} 

	\centering 
	
	\scshape 
	
	\vspace*{\baselineskip} 
	
	
	\rule{\textwidth}{1.6pt}\vspace*{-\baselineskip}\vspace*{2pt} 
	\rule{\textwidth}{0.4pt} 
	
	\vspace{0.75\baselineskip} 
	
	{\LARGE DCNetBench: \\ Scaleable Data Center Network Benchmarking\\} 
	
	\vspace{0.75\baselineskip} 
	
	\rule{\textwidth}{0.4pt}\vspace*{-\baselineskip}\vspace{3.2pt} 
	\rule{\textwidth}{1.6pt} 
	
	\vspace{2\baselineskip} 
	
	
	
	\vspace*{3\baselineskip} 
	
	
	Edited By
	
	\vspace{0.5\baselineskip} 
	
	{\scshape\Large Ke Liu \\ Wanling Gao \\ Chunjie Luo \\ Cheng Huang \\ Chunxin Lan \\ Zhenxing Zhang  \\ Lei Wang \\ Xiwen He \\ Nan Li \\ Jianfeng Zhan }
	
	
	\vspace{0.5\baselineskip} 

	\vfill 
	
	
	\textit{\\BenchCouncil: International Open Benchmark Council\\Chinese Academy of Sciences\\Beijing, China\\http://www.benchcouncil.org} 
	\vspace{5\baselineskip} 

	
	{\large Feb 20, 2023} 

\end{titlepage}



\title{DCNetBench: Scaleable Data Center Network Benchmarking}

\author[1,2]{Ke Liu}
\author[1,3]{Wanling Gao\thanks{Wanling Gao is the corresponding author.}}
\author[1,3]{Chunjie Luo}
\author[1,2]{Cheng Huang}
\author[1]{Chunxin Lan}
\author[4]{Zhenxing Zhang}
\author[1,2,3]{Lei Wang}
\author[1]{Xiwen He}
\author[4]{Nan Li}
\author[1,2,3]{Jianfeng Zhan}

\affil[1]{Research Center for Advanced Computer Systems, State Key Lab of Processors, Institute of Computing Technology, Chinese Academy of Sciences \\ \{liuke2018, gaowanling, luochunjie, lanchuanxin, wanglei\_2011, hexiwen, zhanjianfeng\}@ict.ac.cn}
\affil[2]{University of Chinese Academy of Sciences, huangcheng14@mails.ucas.edu.cn}
\affil[3]{BenchCouncil (International Open Benchmark Council)}
\affil[4]{Huawei, zhang.zx@huawei.com, lee.linan@huawei.com}

\date{Feb 20, 2023}

\maketitle

\begin{abstract}
	Data center networking is the central infrastructure of the modern information society. However, benchmarking them is very challenging as the real-world network traffic is difficult to model, and Internet service giants treat the network traffic as confidential. Several industries  have published a few publicly available network traces. However, these traces are collected from specific data center environments, e.g., applications, network topology, protocols, and hardware devices, and
	thus cannot be scaled to different users, underlying technologies, and varying benchmarking requirements. 
	
	This article argues we should scale different data center applications and environments in  designing, implementing, and evaluating  data center networking benchmarking. We build DCNetBench, the first application-driven data center network benchmarking that can scale to different users, underlying technologies, and varying benchmarking requirements.  The methodology is as follows. We built an emulated system that can simulate networking with different configurations. Then we run applications on the emulated systems to capture the realistic network traffic patterns; we analyze and classify these patterns to model and replay those traces. Finally, we provide an automatic benchmarking framework to support this pipeline. The evaluations on DCNetBench show its  scaleability, effectiveness, and diversity for data center network benchmarking.
	
	
\end{abstract}

\clearpage

\section{Introduction}

Datacenters (DCs) are fundamental infrastructures that occupy a 15.2 Billion  USD switch market and deliver 20.6 zettabytes (ZB) of global network traffic per year by 2021, according to the forecasts of market study report ~\cite{IGR_2021_report_SwitchMarket}, and Cisco ~\cite{Cisco_2018_WhitePaper_DataCenter}.
Hence, how to design, evaluate, and optimize DC network architectures, protocols, network resource management, and hardware devices are of paramount significance in terms of the severe network latency and bandwidth requirements.
Toward this purpose, benchmarks lay the foundation and play vital roles.
However, DC's large-scale deployment, complexity, and confidentiality pose great challenges.

On the one hand, the measurable networking properties depend on the problem definition (applications) and solution instantiation (different implementation techniques)~\cite{zhan2022benchcouncil}. The problem definition, solution instantiation, and measurement are entangled and have complex mutual influences~\cite{zhan2022benchcouncil}. Diverse real-world applications implemented with different programming languages and frameworks, network architectures, topologies, protocols, and the hardware devices like CPU processors and switch chips would largely influence the network traffic patterns. Conversely, different DC network traffic patterns largely impact the networking design decisions and configurations of DC infrastructures like topology, protocol, switch chip, etc. Hence, a benchmark methodology and tool should have the ability to  be scaled to different applications, underlying technologies like network topology, protocols, hardware devices, and varying benchmarking requirements.

For example, taking into account the fact of applications, if you are a search engine or ChatGPT provider, definitely, the network traces generated from a social networking provider does not make sense in evaluating and measuring your DC networking.  Similarly,  for the same application, different underlying technology like network topology, protocols, and hardware devices will significantly impact the generated network traces, and hence have an effect on the measured quantities. 
In a word, the benchmarks should be scaleable to different users and underlying technologies: It not only reflects the different real-world DC network traffic patterns,  but also  the impacts of underlying technologies. Unfortunately, tackling these issues is challenging. For example, DC network traffic is considered a confidential issue and is in the hands of industry giants. 



On the other hand, a DC usually contains thousands of machine nodes and provides billion-level services, requiring a meager response time of several milliseconds or even microseconds.
Thus, it is hard to perform such a large-scale study, whether from the perspective of obtaining vast machine resources or from the perspective of performing the deployment and evaluations on a data center directly.
Furthermore, even though we have overcome these troubles, the highly mixed, multifarious, and complex online or offline data center applications aggravate the difficulties of exploring the mainstream network traffic patterns and further constructing corresponding traces for benchmarking. In this context, the community needs an automatic framework to facilitate large-scale benchmarking (scalable) with applications. Please note the difference between being scaleable and scalable. The former refers to that a benchmark can be scaled to different users, system sizes,  and underlying technologies, while the latter indicates the size can be scaled up and down. The former contains the latter.

To tackle the challenges, state-of-the-art or state-of-the-practice efforts mainly adopt three kinds of methodologies and benchmarks.
The first one is to construct benchmarks according to the standards defined by authoritative institutions. Specifically, the Internet Engineering Task Force (IETF) published a series of RFC documents to determine the evaluation methodology and measurement techniques for network devices in data center~\cite{Bradner_1999_RFC2544_BenchmarkingMethodology, Avramov_2017_rfc8239_DCBenchMethod, Mandeville_2000_rfc2889_SwitchBenchMethod, Bradner_1991_RFC1242_NetworkTerminology}. 
The second one is to use mathematical models to simulate the network traffic patterns. For example, the uniform and Bernoulli models are widely used to simulate the packet patterns~\cite{Papaphilippou_2020_SIGDA_SwitchFPGA, Hassen_2017_ICC_ClosSwitch, He_2011_ICC_trafficAlgorithm, Li_2002_IEEE_dualRoundRobin}.
The problem is either RFC documents or mathematical models are far from real-world network traffic characteristics. Thus, their evaluation results may not reflect the actual design requirements.
The third category uses realistic traces published by industry giants to achieve reality and validity. For example, Microsoft, Google, Facebook, etc., have published a few publicly available network traces or analyses on large-scale production clusters~\cite{Srikanth_2009_SIGCOMM_DCTraffic, Benson_2010_SIGCOMM_DCTraffic, Roy_2015_SIGCOMM_DCNetwork}. 
However, these traces are collected from the specific data center environment and specific applications and thus cannot be scaled to different users, underlying technologies, and varying benchmarking requirements. In addition, several of them only disclose limited statistical characteristics obtained from the traffic data, which is insufficient to see the whole picture. Some are not publicly available. Even though there have several open-sourced network traces, they provide either short-term traffic information or limited statistical fields like IP.  These data are hard to be used directly for benchmarking and cannot form the basis for traffic generation either.

\begin{figure}[htb]
	\centering
        \includegraphics[width=0.6\linewidth]{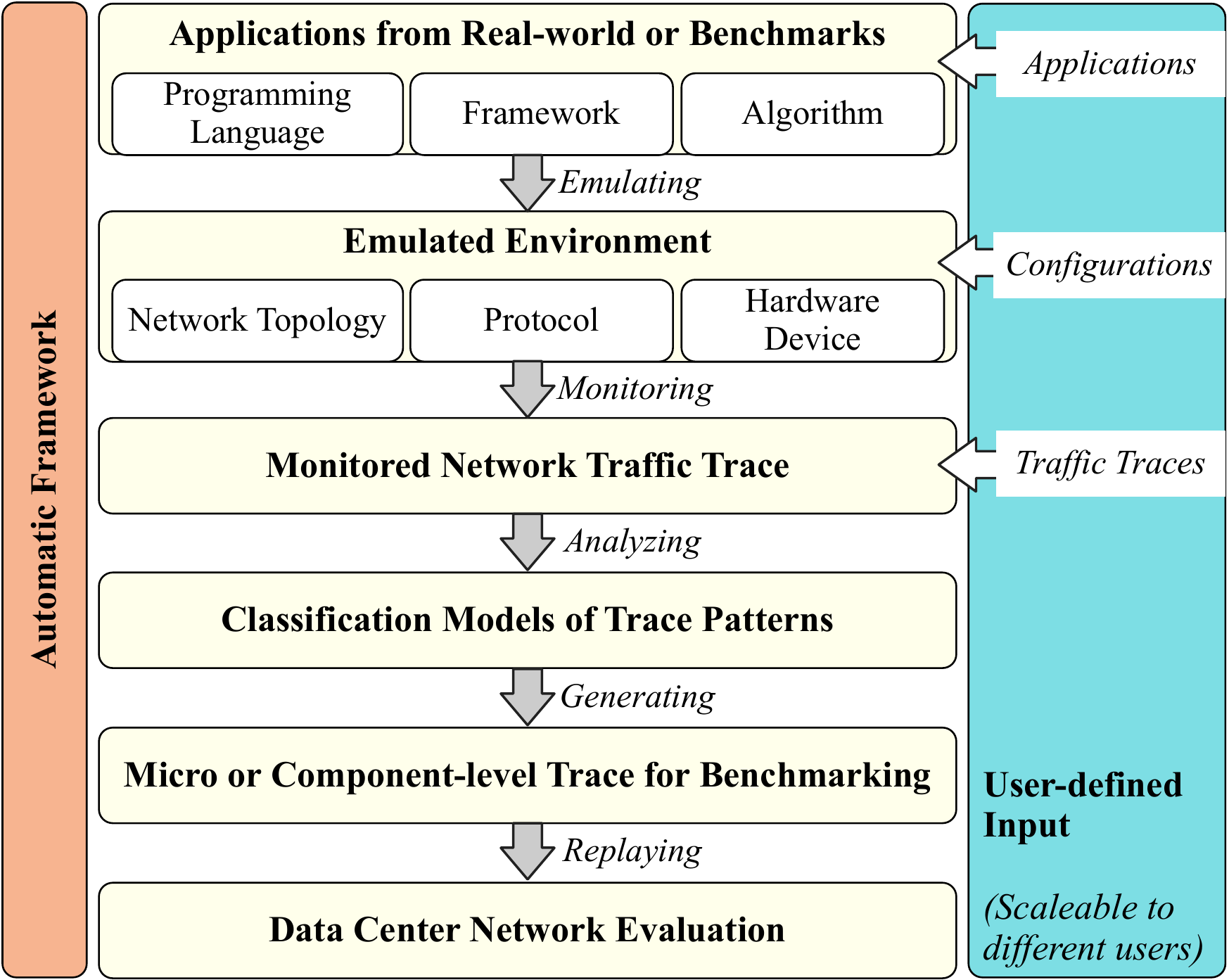}
	\caption{DCNetBench Methodology.}
	\label{1-metho}
\end{figure}

In this paper, we propose a scaleable DC networking benchmarking methodology, framework, and tool -- DCNetBench, as shown in Fig.~\ref{1-metho}. We achieve being scaleable and realistic by receiving user-defined configurations, deploying real-world applications on an emulated system, monitoring the network traffic of an actual running, and constructing micro traces and component-level traces that maintain the monitored patterns. 
Note that a micro trace reflects the representative network traffic extracted from the whole application execution with a single pattern, like a burst. A component-level trace means the network traffic of a whole application execution with a combination of different patterns.
The DCNetBench methodology and tool can be scaleable to different users,  real-world applications, network topologies, protocols, and hardware devices, which means it supports generating corresponding traces when using a different application, network topology, protocol, or switch type, or emulating on a different hardware platform like X86 or ARM.  
The comparison of DCNetBench with the related work is shown in Table~\ref{table_Benchmarking_Method_summary}.
Based on the methodology, we provide an automatic benchmarking framework that supports automatically generating traces and replaying them for evaluation. It includes four loosely coupled modules: environment deployer based on MaxiNet~\cite{Philip_2014_IFIP_MaxiNet}, trace generator, trace evaluator, and metric reporter. 
They support both individual and collective deployment and execution with a simple configuration. 
Using the typical applications in BigDataBench~\cite{WangLei_2014_HPCA_BigDataBench} and AIBench~\cite{WanlingGao_2018_Springer_AIBench,tang2021aibench} as an example, we construct twelve micro traces and four component-level traces for network benchmarking.
Users can generate their own traces for evaluation by defining different configurations. The input could be applications in widely-used benchmarks, real-world scenarios, or network traces.

\begin{table*}[htb]
	\caption{Benchmarking Methodology Comparison.}
	\label{table_Benchmarking_Method_summary}
		\resizebox{1\linewidth}{!} {
			\small
		\begin{tabular}{cccccc}
			\toprule
			\textbf{Methodology} & \textbf{Real-world Application}  & \textbf{Network Topology}  &  \textbf{Protocol}  & \textbf{Hardware Devices}  & \textbf{Automatic Framework}                                    \\
			\midrule
			RFC Documents~\cite{Bradner_1991_RFC1242_NetworkTerminology,Bradner_1999_RFC2544_BenchmarkingMethodology,Mandeville_2000_rfc2889_SwitchBenchMethod,Avramov_2017_rfc8239_DCBenchMethod}                         & \ding{56}                    &  \ding{56}                        & \ding{56}                           & \ding{56} &   \ding{56} \\
			Mathematical models~\cite{ZefuDai_2012_SwitchFPGA,Andrew_2014_ANCS_SwitchFPGA,Hassen_2017_ICC_ClosSwitch,Papaphilippou_2020_SIGDA_SwitchFPGA}                   & \ding{56}                    & \ding{56}      & \ding{56}                           & \ding{56}         & \ding{56}                 \\
			Network Traces~\cite{Benson_2010_SIGCOMM_DCTraffic,Facebook_web_NetworkTraffic,Yahoo_dataset_traffic,Fontugne_2010_coNext_MAWILab}                & Fixed                            & Fixed                        & Fixed                          & Fixed &   \ding{56}                        \\
			DCNetBench                   & Scaleable                         & Scaleable     & Scaleable & Scaleable   & \ding{52}                          \\
			\bottomrule
		\end{tabular}
	}
\end{table*}

The rest of this paper is organized as follows:
Section 2 introduces the motivation; section 3 shows the related work. Section 4 illustrates the design and implementation.
Section 5 presents our evaluations, and section 6 draws a conclusion.

\section{Motivation}


The network traffic patterns determine the software and hardware design decisions on DC networking. How to model real-world network traffic is a critical problem.
The state-of-the-art and state-of-the-practice usually adopt RFC specifications, mathematical models, or traffic traces from DC giants for network benchmarking.
In this section, we illustrate why we need a new benchmarking methodology.

\subsection{Scaleability is Essential for Network Benchmarking}

\begin{figure}[htbp]
	\centering
	\begin{minipage}[c]{0.48\textwidth}
		\subfloat[Traffic Matrix flowing through A Core Switch using Three-tier Tree]{
			\includegraphics[width = 0.44\linewidth]{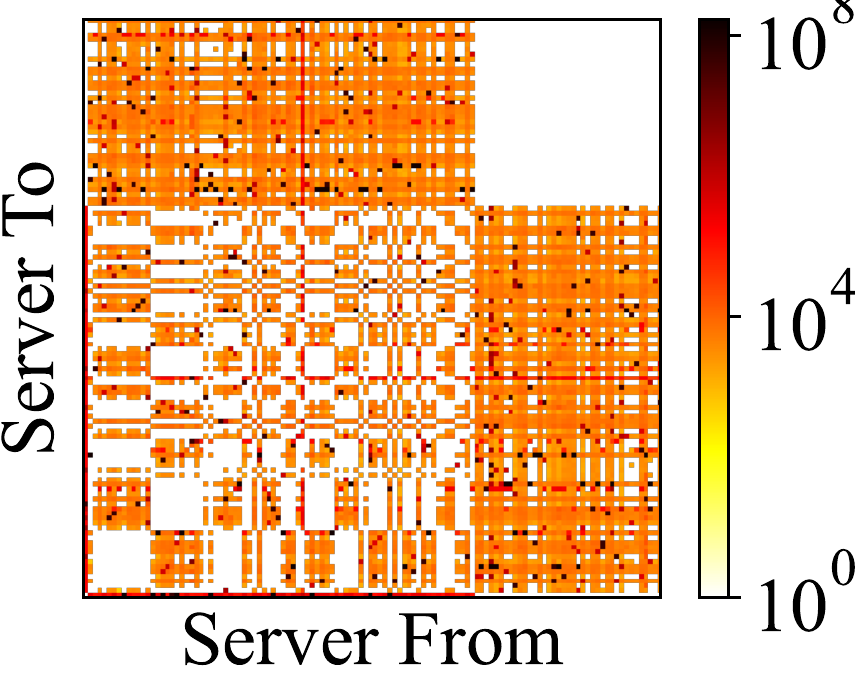}
			\label{figure_tree_coreswitch_h_MatrixMult_traffic_matrix}
		} \quad
		\subfloat[Traffic Matrix flowing through A Core Switch using Spine-leaf] {
			\includegraphics[width = 0.44\linewidth]{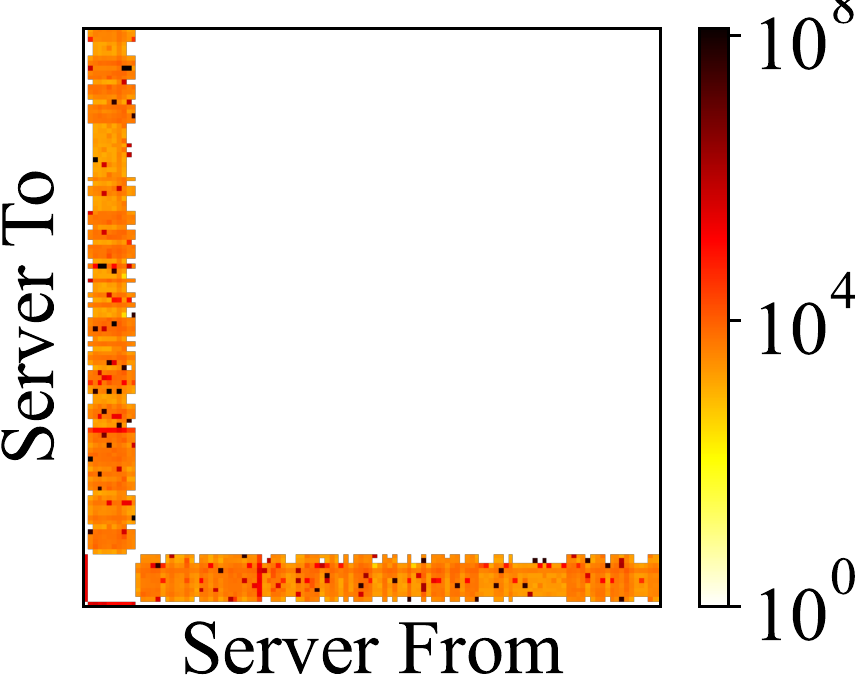}
			\label{figure_spineleaf_coreswitch_h_MatrixMult_traffic_matrix}
		}
		\caption{Network Traffic using Different Topologies.}
		\label{figure_different_topo_coreswitch_h_MatrixMult_traffic_matrix}
	\end{minipage}
	\quad
	\begin{minipage}[c]{0.48\textwidth}
			\centering
		\subfloat[Traffic Matrix of Hadoop MatrixMult]{
			\includegraphics[width = 0.45\linewidth]{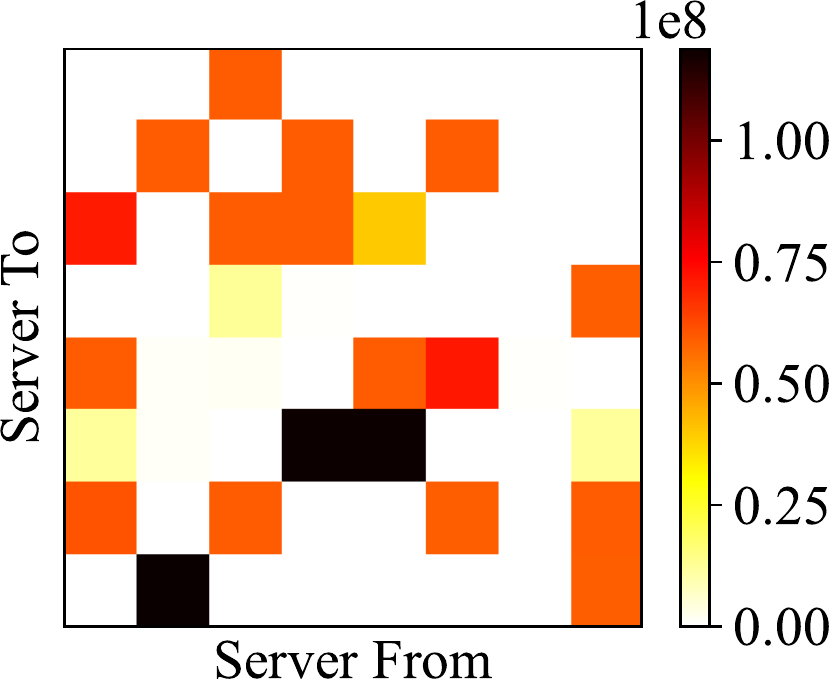}
			\label{figure_validity_CDF_h_grep_2GB_1}
		} \quad
		\subfloat[Traffic Matrix of Spark PageRank]{
			\includegraphics[width = 0.43\linewidth]{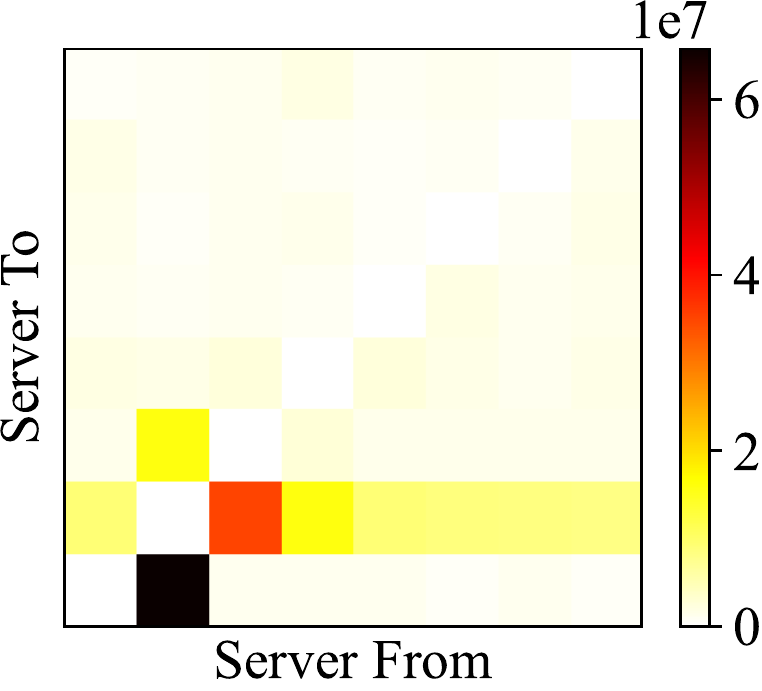}
			\label{figure_traffic_matrix_s_PageRank_20_1}
		} \\
		\caption{Network Traffic using BigDataBench~\cite{WangLei_2014_HPCA_BigDataBench}.}
		\label{figure_traffic_matrix_real_data_center}
	\end{minipage}
\end{figure}

Using applications and DC environment as examples, we conduct experiments to motivate the necessity of scaleable benchmarking. We use the traffic matrix for comparison, an important metric that indicates the amount of data transmission among server nodes~\cite{telkamp2005best}.
Fig.~\ref{figure_different_topo_coreswitch_h_MatrixMult_traffic_matrix} shows the traffic matrix statistics flowing through a core switch using two network topologies: Spine-leaf and Three-tier. They run the same MatrixMult (i.e., matrix multiplication) application. We find that their patterns have distinct differences, indicating the network topology and devices have impacts on the network traffic patterns.
For different applications, we compare two representative applications from a widely-user big data benchmark suite BigDataBench~\cite{WangLei_2014_HPCA_BigDataBench}, as shown in Fig.~\ref{figure_traffic_matrix_real_data_center}.
We find that the traffic matrices of Hadoop MatrixMult and Spark PageRank have distinct differences, indicating the  impacts of algorithms and frameworks on network traffic patterns.
Thus, a scaleable benchmarking that can capture these impacts and generate corresponding traces is essential. 

Even though several traffic traces from DC giants achieve the reality, however, these traces are collected from specific DC environments and applications and cannot be scaled to different users, underlying technologies, and varying benchmarking requirements. On the other hand, 
most of them only provide partial information and have limitations for replaying.  Several of them only report their statistics without the open access traces, and we are hard to reproduce the network traffic according to the characterization data. Table ~\ref{table_motivation_exists_trace_info} shows the overview of realistic traces.
Facebook provides the open access network trace and has enough information for replaying. However, 
its trace is outdated with the latest update in 2017, and is hard to fulfill the scaleable requirements.

	\begin{table*}[]	
		\caption{The Overview of the Traffic Traces from DC Giants.}
		\label{table_motivation_exists_trace_info}
			\resizebox{1.0\linewidth}{!} {
				\begin{threeparttable}
								\begin{tabular}{cccccc}
									\toprule
									\textbf{Source} &     \textbf{Scenario} & \textbf{Latest update} & \textbf{Duration}  & \textbf{Open access}  & \textbf{Enough fields for generation} \\
									\midrule
									Facebook~\cite{Roy_2015_SIGCOMM_DCNetwork}    & Social Media    & 2017   & 1 day   & \ding{52} & \ding{52} \\
									Microsoft ~\cite{Alizadeh_2010_SIGCOMM_DCTCP} & Web Search              & 2010   & 1 month   & \ding{56}    & -  \\
									Microsoft ~\cite{Benson_2010_SIGCOMM_DCTraffic} & Education, Commercial & 2010 & $\sim$ 10 days & \ding{56}  & -\\
									Microsoft ~\cite{Srikanth_2009_SIGCOMM_DCTraffic, Greenberg_2009_SIGCOMM_VL2} & Big Data & 2009  & 2 month  & \ding{56} & -  \\
									Yahoo~\cite{Yahoo_dataset_traffic}            & Web search       & 2007   & 1 day    & \ding{52}         & \ding{56}    \\
									

	\bottomrule
		\end{tabular}
	\end{threeparttable}
}
\end{table*}

	\subsection{Do RFCs and Mathematical Models Capture Realistic Network Patterns?} \label{rfc_math_model_weakness}
	As two widely-used benchmarking methodologies, do RFCs and mathematical models capture the realistic network patterns of DC applications? Our answer is NO!

\begin{figure}[htbp]
	\centering
	\begin{minipage}[c]{0.48\textwidth}
		\subfloat[Traffic Matrix of All-to-one]{
			\includegraphics[width = 0.44\linewidth]{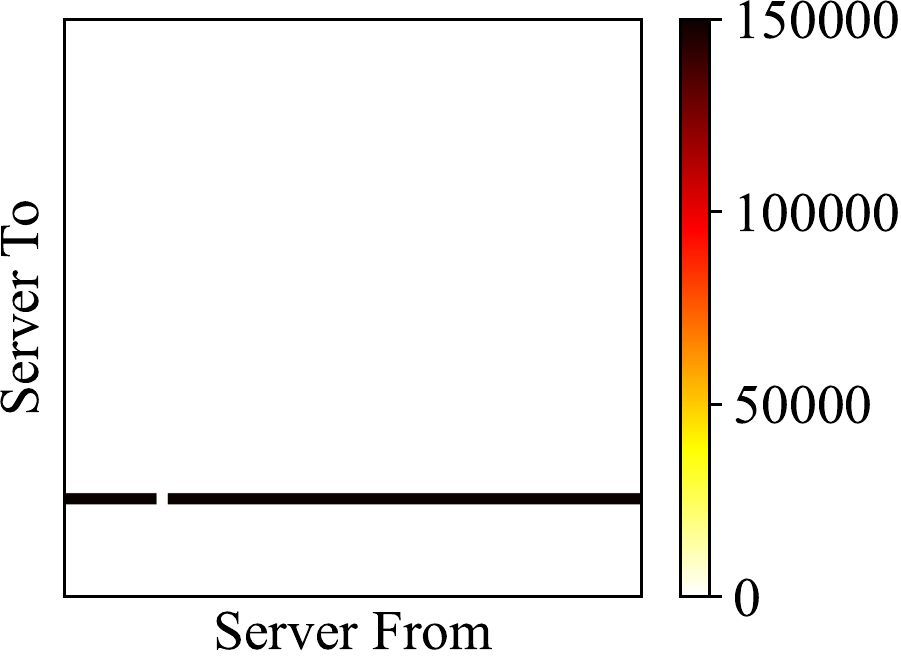}
			\label{figure_traffic_matrix_RFC_all2one_50x10000}
		} \quad
		\subfloat[Traffic Matrix of One-to-all] {
			\includegraphics[width = 0.44\linewidth]{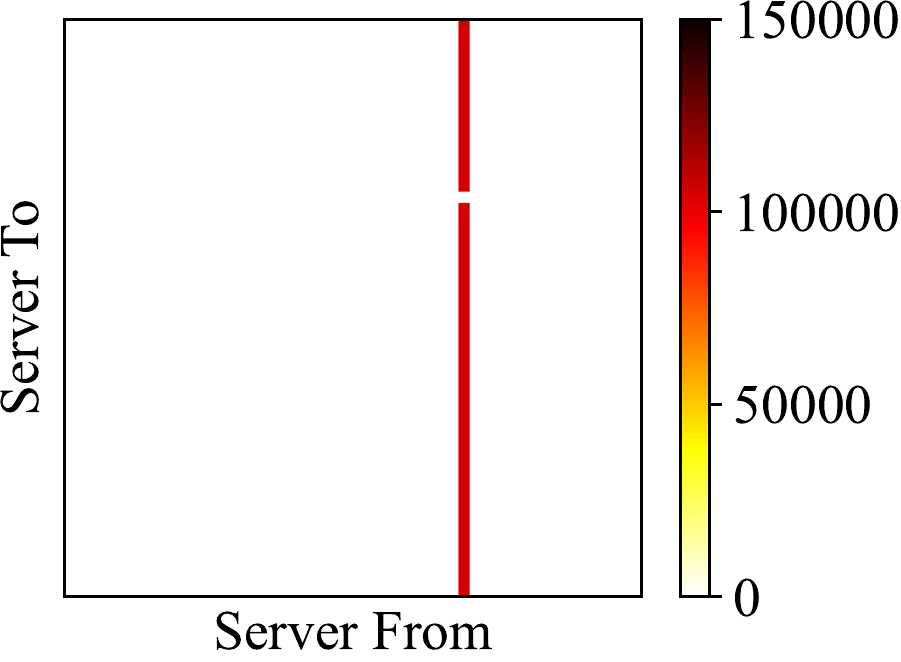}
			\label{figure_traffic_matrix_RFC_one2all_50x10000}
		}
		\caption{Network Traffic using RFC 8239 ~\cite{Avramov_2017_rfc8239_DCBenchMethod}.}
		\label{figure_traffic_matrix_rfc}
	\end{minipage}
	\quad
	\begin{minipage}[c]{0.48\textwidth}
		\subfloat[Traffic Matrix of Uniform Pattern ~\cite{ZefuDai_2012_SwitchFPGA}]{
			\includegraphics[width = 0.44\linewidth]{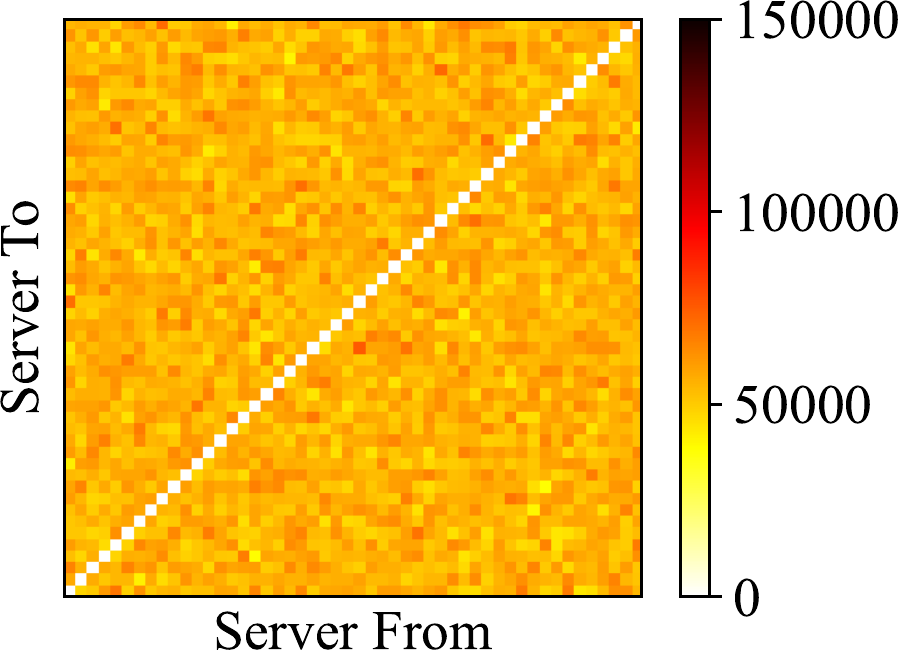}
			\label{figure_traffic_matrix_uniform_50x10000}
		} \quad
			\subfloat[Traffic Matrix of Nonuniform Pattern~\cite{Andrew_2014_ANCS_SwitchFPGA}] {
				\includegraphics[width = 0.44\linewidth]{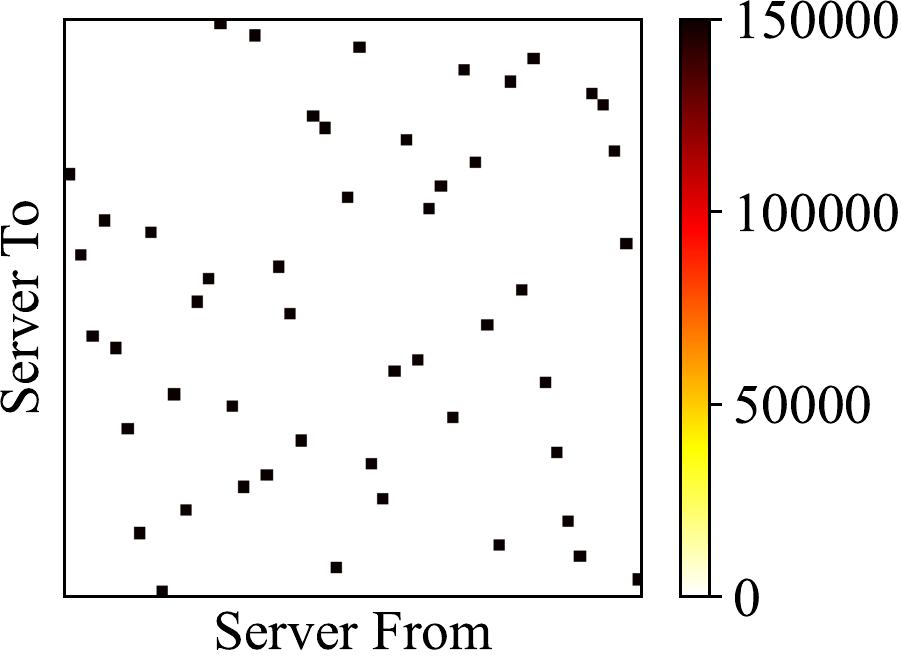}
				\label{figure_traffic_matrix_permutation_digit_shuffle_50x10000}
			}
			\caption{Network Traffic using Mathematical Models.}
			\label{figure_traffic_matrix_mathmodel}		
	\end{minipage}	
\end{figure}

	We compare the network traffic of RFC (Fig.  ~\ref{figure_traffic_matrix_rfc}) and mathematical models (Fig. ~\ref{figure_traffic_matrix_mathmodel}) with the applications in Fig.~\ref{figure_traffic_matrix_real_data_center}.
	We find that their traffic matrices have significant distinctions. 
	Specifically, the traffic matrix using RFC 8239 -- the latest related version of the RFC document -- has stable packet sizes and only a tiny fraction of the total pairs of server nodes have data transmission. The traffic matrices using two widely used mathematical models -- uniform (Fig. ~\ref{figure_traffic_matrix_mathmodel}(a)) and nonuniform (Fig. ~\ref{figure_traffic_matrix_mathmodel}(b)) models --  reflect either uniform or punctate patterns.  Both of them fail to capture the realistic network patterns shown in Fig. ~\ref{figure_traffic_matrix_real_data_center}.

\subsection{The Effectiveness of Emulation}

Data center usually contains hundreds or thousands or even more physical machines, 
which is difficult to obtain, maintain, and use for evaluation directly.
In this condition, DC network emulators are proposed to emulate a large-scale DC environment on small-scale physical machine resources. 


Mininet is a widely-used network emulator that supports "rapidly prototyping large networks on the constrained resources of a single laptop"~\cite{Bob_SIGCOMM_2010_Mininet}. 
MaxiNet~\cite{Philip_2014_IFIP_MaxiNet} is a distributed network emulator based on Mininet that supports network emulation on a physical cluster with multiple server nodes. It uses a software switch -- Open vSwitch (OVS)~\cite{ovs_website} to simulate hardware switches' functionality and containers to simulate physical nodes. 
These emulators enable defining the network architecture according to realistic topologies and adopt realistic network protocols. Thus theoretically, they have the ability to emulate the realistic DC architecture and network environment.
Meanwhile, their evaluations show the effectiveness of the proposed systems for large-scale emulations~\cite{Philip_2014_IFIP_MaxiNet, Bob_SIGCOMM_2010_Mininet}. 
Our experiments validate this point and show that emulated networking behaviors can represent realistic behaviors with high confidence.


\begin{figure}[htb]
	\centering
	\includegraphics[width=0.5\linewidth]{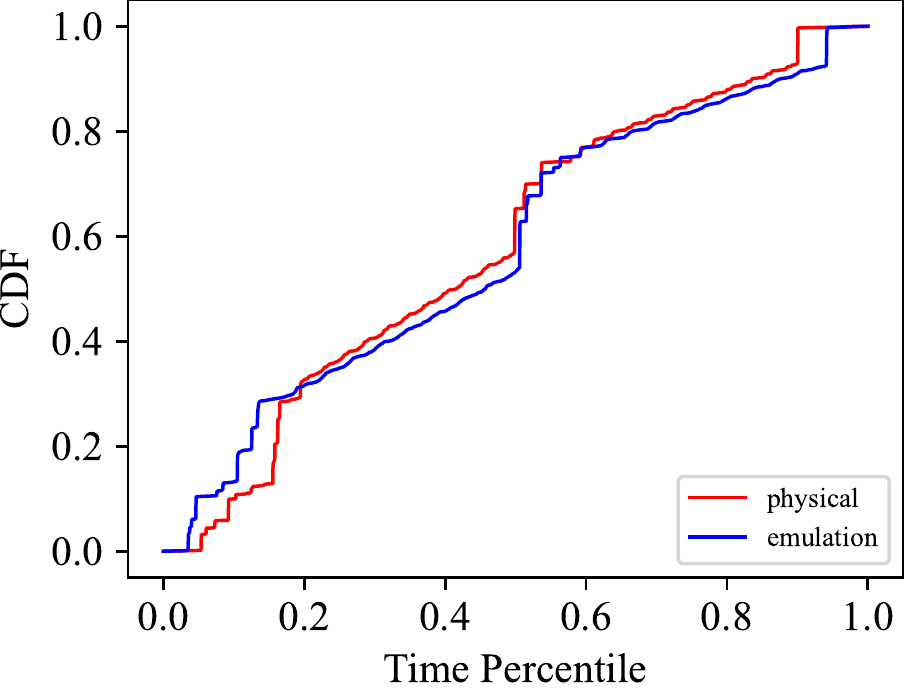}
	\caption{Network Traffic CDF of Hadoop Grep.}
	\label{figure_validity_CDF_h_grep_2GB_1_12point}
\end{figure}

Figure~\ref{figure_validity_CDF_h_grep_2GB_1_12point} shows the comparison of network traffic on an 8-node physical cluster and an 8-node emulated cluster. We use a typical big data application implemented with Hadoop framework~\cite{Shvachko_2010_IEEE_Hadoop} -- Hadoop Grep. We compare their similarities using the cumulative distribution function (CDF) as a metric example. From Figure~\ref{figure_validity_CDF_h_grep_2GB_1_12point},  we find that their CDF curves are nearly the same, which means they reflect consistent network behaviors. The detailed evaluations using other applications and metrics are shown in Section~\ref{eva-ver}. 

Above all, we argue that modeling real-world network traffic through emulation is practical.
It supports emulating a large-scale network architecture using limited physical resources while guaranteeing network traffic's reality. To the best of our knowledge, none of the previous works has adopted such an approach for benchmarking.


\section{Related work}

Several benchmarks and standards have been proposed for DC network evaluation and optimization. 

The Internet Engineering Task Force (IETF) published a series of RFC (Request for Comment) documents to define the network benchmarking methodology or traffic specifications. For example, RFC 1242~\cite{Bradner_1991_RFC1242_NetworkTerminology} defined the benchmarking terminology for network interconnection devices in 1991. RFC 2544~\cite{Bradner_1999_RFC2544_BenchmarkingMethodology} specified benchmarking methodology for network interconnect devices in 1999. RFC 2889~\cite{Mandeville_2000_rfc2889_SwitchBenchMethod} standardized the benchmarking methodology for LAN switching devices in 2000. RFC 8239~\cite{Avramov_2017_rfc8239_DCBenchMethod} regulated the data center benchmarking methodology in 2017. 
RFCs have been widely used in industry. For example, Huawei used RFC 2544 to evaluate their CloudEngine 12800 series DC switches~\cite{Miercom_2012_report_CE12800Switch}. Cisco used RFC 2544, 2889, and other documents to assess the Nexus 9516 switch~\cite{Cisco_2015_report_CiscoNexus9516}. Based on the RFC documents, the Tolly report library compared the performance of Mellanox Spectrum switch and Broadcom StrataXGS Tomahawk switch~\cite{Tolly_2016_report_SwitchBench}.

Another attempt is to benchmark and evaluate the networking performance using mathematical models. 
Specifically, they use different mathematical models to simulate various aspects of network patterns, e.g., the packet arrival pattern, packet size pattern, and packet transmission pattern of every pair of source and destination serve nodes.
Dai et al.~\cite{ZefuDai_2012_SwitchFPGA} adopted the Bernoulli model to simulate the packet arrival pattern, with the packet size randomly selected from 32 bytes or 512 bytes.
Bitar et al.~\cite{Andrew_2014_ANCS_SwitchFPGA} evaluated an ethernet switch with an NoC-enhanced FPGA and simulated several network patterns. They adopted a Bernoulli model for packet arrival pattern, a permutation model for source-destination packet transmission pattern, and a normal distribution selecting from ten groups ranging from 64 bytes to 1504 bytes for packet size pattern.
Hassen et al.~\cite{Hassen_2017_ICC_ClosSwitch} adopted uniform Bernoulli, hotspot Bernoulli, and diagonal traffic models to evaluate the high-capacity switches.
Papaphilippou et al.~\cite{Papaphilippou_2020_SIGDA_SwitchFPGA} adopted uniform Bernoulli, uniform burst (Markov chain), and nonuniform traffic models to evaluate a high-performance FPGA network switch architecture.

Several DC giants published their historical network traffic traces or analysis, which can be used for traffic generation and benchmarking. 
Microsoft~\cite{Benson_2010_SIGCOMM_DCTraffic, Srikanth_2009_SIGCOMM_DCTraffic,Greenberg_2009_SIGCOMM_VL2,Alizadeh_2010_SIGCOMM_DCTCP,YiboZhu_2015_SIGCOMM_RDMA,MicrosoftAzure_github_ResourceTrarffic} released network traces and performed analysis, however, it lacks important information like packet size for traffic generation.
Facebook~\cite{Facebook_web_NetworkTraffic} released a network trace with a 24-hour duration.
Google~\cite{Google_github_ResourceTrace} and Alibaba~\cite{Alibaba_github_ResourceTraffic} released their DC traces, however, they are mainly resource utilization information and have little network information.
Yahoo~\cite{Yahoo_dataset_traffic} published network traces that were collected on three border routers.
UNIBS~\cite{UNIBS_web_SecurityTraffic} published network traces that were collected on the edge router of the campus network. 
MAWI~\cite{MAWI_web_BackboneTraffic}, CAIDA~\cite{caida_webDB_BackboneTrace}, and MAWLab ~\cite{Fontugne_2010_coNext_MAWILab} published a series of backbone traffic traces. These traces from edge router or backbone traffic traces only include the traffic transmission among multiple DCs instead of the network information within a DC.

Several traffic generation tools are provided for network evaluation. 
Adeleke et al.~\cite{Adeleke_2022_CSUR_trafficGenerateSurvey}  surveyed 92 traffic generators and concluded that more than 25\% of them, e.g., netperf ~\cite{Netperf_Benchmark} and iperf ~\cite{ChungHsing_1998_document_iperf,SourceForge_iperf2,LBNL_2014_github_iperf3}, only support a set of fixed parameters to perform evaluations. 
Scapy ~\cite{Scapy_PacketGen} and Pketgen-dpkt~\cite{Wiles_2010_github_PktgenDPDK} are script-driven traffic generators that allow dynamically modify~\cite{Adeleke_2022_CSUR_trafficGenerateSurvey}.
Heeperf~\cite{Mosberger_1998_ACM_httperf} generates traffic traces based on HTTP protocol.
Harpoon~\cite{Sommers_2004_ACM_Harpoon} is a tool that supports modeling an input trace file and generating similar network traffic.
HP Labs~\cite{Mahadevan_2009_ICRN_SwitchBenchmark}  constructed a generator that generated broadcast traffic for energy consumption evaluation.

All these efforts fail to guarantee the reality of network patterns and are hard to be scaled to different users, underlying technologies, and varying benchmarking requirements. Even though the real-world traffic traces remain realistic patterns, however, we are hard to reproduce the reality in benchmarking due to the limited information. Additionally, the traces are collected from the specific DC environment and configurations. Above all, we argue that we urgently need a new methodology for DC network benchmarking that is application-driven and meanwhile scaleable.

\section{Design and Implementation}

This section presents DCNetBench, a scaleable application-driven methodology, framework, and tool for DC network benchmarking. We first introduce the scaleable methodology (\textbf{Subsection~\ref{design-metho}}). After that, we illustrate the design and implementation of the automatic benchmarking framework 
(\textbf{Subsection~\ref{design-framework}}). Finally, based on the framework, we construct micro and component-level traces for network benchmarking (\textbf{Subsection~\ref{design-dcnet}}). 


\subsection{Scaleable Benchmarking Methodology}~\label{design-metho}

We propose a scaleable application-driven benchmarking methodology to model the realistic network traffic (Fig.~\ref{1-metho}). 
The methodology captures the mutual effects of applications and DC environment on network traffic and can scale to different users, underlying technologies, and
varying benchmarking requirements. 
First, for an application from real-world or benchmarks (e.g., BigDataBench~\cite{WangLei_2014_HPCA_BigDataBench}) implemented with a programming language and framework, we deploy it on an emulated environment with a configuration of network topology, protocol, and software switch. Note that the emulated environment supports being deployed on different physical devices like X86 or ARM clusters. Second, we run the application and monitor its runtime networking performance. After that, we obtain the traffic trace, analyze the patterns, and perform classifications. Then we generate the micro and component-level network traces according to the classification models. A micro trace contains a single pattern like a burst, while the component one contains a combination of different patterns. Based on the generated traces, we replay them for DC network benchmarking. Finally, we provide an automatic framework to support this pipeline. 

The methodology can be scaled to different users through user-defined input. The input can be an application, a traffic trace, or an environment configuration that specifies the network topology, protocol, or hardware device. The DCNetBench will generate corresponding micro and component-level traces for benchmarking.


\subsection{Automatic Benchmarking Framework}~\label{design-framework}

\begin{figure*}
	\centering
            \includegraphics[width=1.0\linewidth]{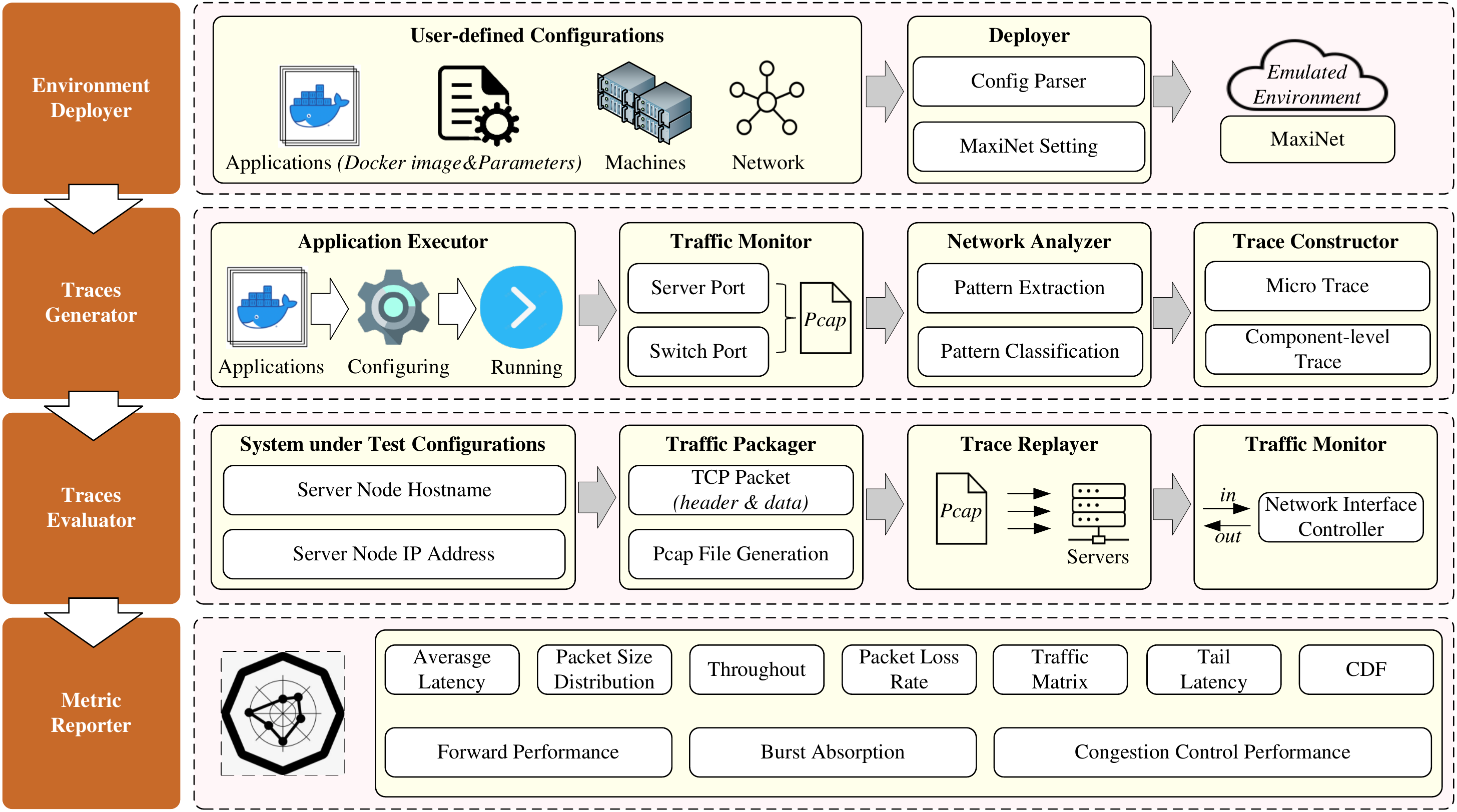}
	\caption{Automatic Benchmarking Framework Architecture.}
	\label{figure_Framework_architecture}
\end{figure*}

As shown in Fig.~\ref{figure_Framework_architecture}, the automatic benchmarking framework includes four loosely coupled modules: environment deployer, trace generator, trace evaluator, and metric reporter. They support both individual and collective deployment and execution with a simple configuration. 

\subsubsection{Environment Deployer} The environment deployer is responsible for emulating a large-scale DC environment under limited physical resources, which is built based on  MaxiNet~\cite{Philip_2014_IFIP_MaxiNet}. Each emulated node is a docker container, started from a user-defined docker image with user-concerned applications.
The users need to input the machine configurations, network configurations, and application configurations. Among them, the machine configurations specify the physical cluster information that will be used for emulation, including the IP address, CPU cores, and memory. The network configurations include the DC network topology and the number of switches and server nodes. 
Application configurations provide a docker image file that contains the user-specified applications and the corresponding execution information including the application name, file directory, execution parameters, and input data. 
We have provided a docker image that contains mainstream DC applications like big data implemented with Hadoop and Spark, AI implemented with TensorFlow and PyTorch by default. Users can use it directly to generate corresponding network traces with different network configurations.

\subsubsection{Trace Generator} The trace generator is responsible for running the user-specified application, monitoring and analyzing the network traffic,  and generating micro and component-level traces. It includes four submodules: application executor, runtime network traffic monitor, network analyzer, and trace constructor. 

The application executor is to execute the specified application according to the configurations in the environment deployer. 
The traffic monitor is to monitor the network transmission among different server nodes.
We use TCPDUMP~\cite{tcpdump_website} to obtain the traffic information by monitoring the ingress traffic of all the virtual ports of emulated switches and server nodes.
TCPDUMP will output the binary logs in a pcap format, including the Ethernet frame header and tail, IP header, and TCP header. 
The network analyzer is to read the pcap file generated by TCPDUMP and analyzes the network information. It first reads the header of every data packet and extracts the critical information like timestamp, the total length of the data packet (Ethernet frame), the length of the TCP payload, the source IP, source port, destination IP, and destination port.
Then the analyzer fuses the network information of all the monitoring files and outputs four-tuple traces: \{timestamp, packet size, ingress port of the switch, egress 
port of switch\}. For these four-tuple traces, the analyzer performs two-level classifications. The first classification is to classify the analyzed traffic traces according to the DC architectures and switch types. The second classification further classifies them according to the network patterns, like stable or burst patterns, the cumulative distribution function (CDF), and packet size distribution. Note that the user can specify the time interval for statistics. We use fifty milliseconds by default. 
The trace constructor is to construct the micro and component-level traces according to the classification models generated by the network analyzer. It chooses the traces from different categories and intercepts a fragment containing the corresponding category's representative patterns. 

\subsubsection{Trace Evaluator} The trace evaluator is to read the traffic trace from the trace generator, packages it as a whole TCP packet with TCP header and data, replays the traffic, and monitors the networking performance.
It includes three submodules: trace packager, trace replayer, and traffic monitor.
To achieve automatic evaluation, the user needs to specify the system under test (SUT) configurations, including the server nodes' hostname and IP address.

Trace packager aims to package the four-tuple trace \{timestamp, packet size, ingress port of the switch, egress 
port of switch\} as a whole TCP packet and generate each pcap file for every server node. 
Trace replayer is to distribute the pcap file to every server node and replay the trace according to the timestamp information. 
The traffic monitor also uses TCPDUMP~\cite{tcpdump_website} to obtain the traffic by monitoring the ingress and egress traffic of the network interface controller on the virtual or physical server nodes.
Note that since the traffic may be insufficient for the SUT and further incur underutilization, the trace evaluator would have a multi-round evaluation by amplifying the traffic multiple times according to the results of the Metric Reporter module.

\subsubsection{Metric Reporter} The metric reporter analyzes the network traffic from the trace evaluator and reports the metrics. Currently, it supports a wide spectrum of metrics including latency, throughput, packet loss rate, etc.

\subsection{Micro and Component-level Traces}~\label{design-dcnet}

Based on the automatic benchmarking framework, we construct twelve micro and four component-level traces. We first identify the representative real-world data center scenarios and use the applications from a widely-used big data benchmark suite -- BigDataBench~\cite{WangLei_2014_HPCA_BigDataBench} and an AI benchmark suite -- AIBench~\cite{WanlingGao_2018_Springer_AIBench,tang2021aibench}.
To cover the application representativeness and diversity, we select six applications with different computation complexities, different frameworks, and different input data. Specifically, The six applications are Grep, WordCount, Random Sampling (RandSample), PageRank, Matrix Multiplication (MatrixMult) applications implemented with Hadoop and Spark framework, and an image classification application implemented with PyTorch framework.
All applications are shown in Table~\ref{implement_all_workloads_info}.

\begin{figure}[htb]
	\centering
	\subfloat[Three-tier tree]{
		\includegraphics[width = 0.5\linewidth]{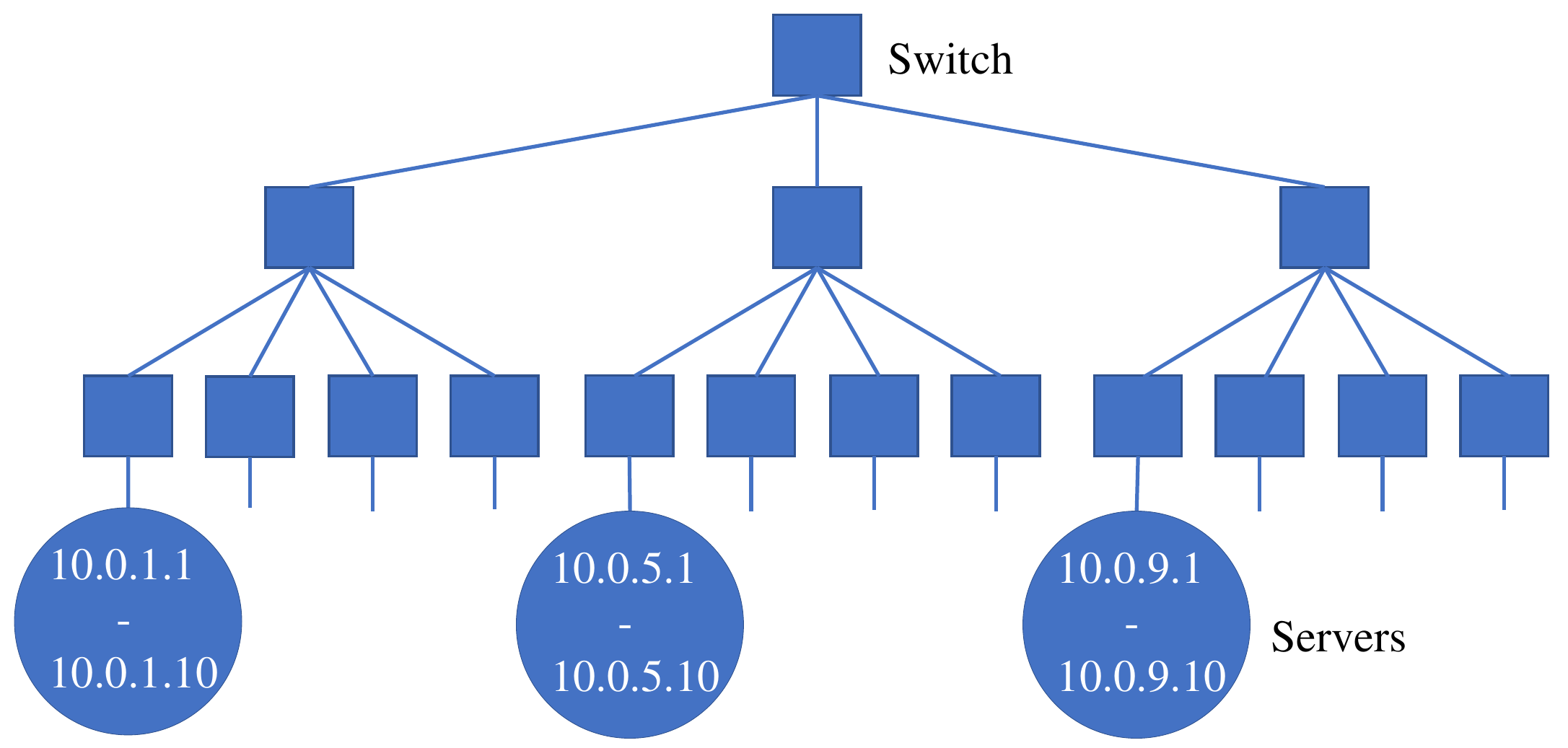}
		\label{figure_implement_topo_simple_tree}
	} \\
	\subfloat[Fat-tree]{
		\includegraphics[width = 0.5\linewidth]{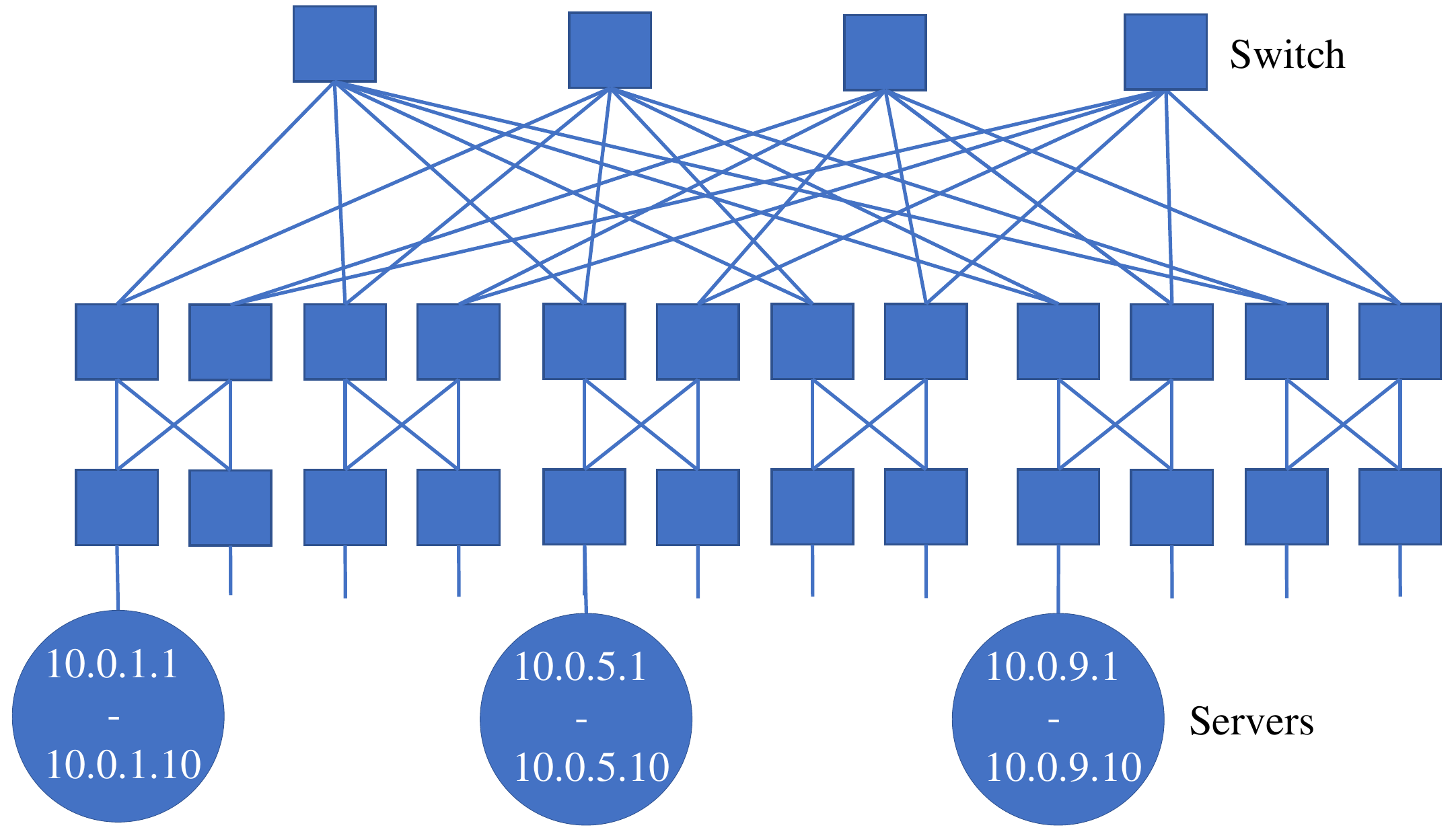}
		\label{figure_implement_topo_fat_tree}
	} \\
	\subfloat[Spine-leaf]{
		\includegraphics[width = 0.5\linewidth]{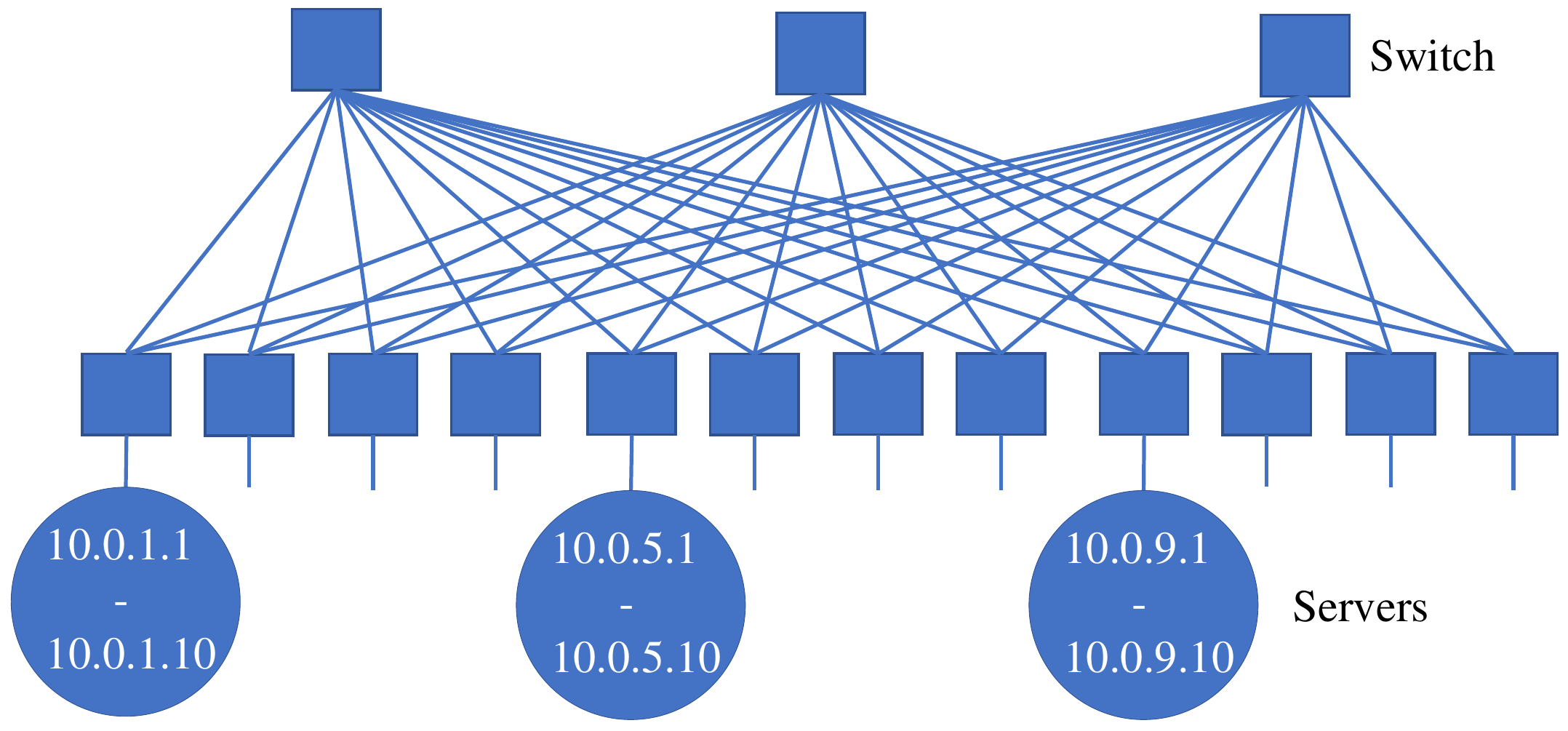}
		\label{figure_implement_topo_spine_leaf}
	} \\
	
	\caption{The Emulated 120-node Cluster using Three Network Topologies.}
	\label{figure_implement_NetworkTopo}
\end{figure}


\begin{table}[htpb]
	\centering
	\caption{Representative DC Applications.}
	\label{implement_all_workloads_info}
		\footnotesize
		\begin{tabular}{cccc}
				\toprule
				\textbf{Scenario}                 & \textbf{Application} & \textbf{Framework} & \textbf{Data}     \\
				\midrule
				\multirow{5}{*}{Big Data} & Grep        & Hadoop, Spark    & Text     \\
				& WordCount   & Hadoop, Spark    & Text     \\
				& RandSample  & Hadoop, Spark    & Text     \\
				& PageRank    & Hadoop, Spark    & Graph     \\
				& MatrixMult  & Hadoop, Spark    & Matrices \\
				\midrule
				AI                       & Image Classification   & PyTorch   & Image    \\  
				\bottomrule
			\end{tabular}
	\end{table}

	For the trace generation, we consider three kinds of network topology: three-tier tree, fat-tree, and spine-leaf architectures, as shown in Fig.~\ref{figure_implement_NetworkTopo}.
	The three-tier tree topology is a classical network architecture and is also called the hierarchical inter-networking model. 
	The Top of Rack (ToR) switches uplink to the aggregation switches and the aggregation switches uplink to the core switches. The core switches are connected to border routers. The specific topology is shown in Fig.~\ref{figure_implement_NetworkTopo}(a).
	Fat-tree~\cite{Mohammad_2008_SIGCOMM_FatTree} is an improved topology to solve the problem of high oversubscription incurred by the three-tier tree, by which the users can deploy cheap and low-speed switches instead of high-speed core switches. 
	Thus, Fat-tree is a widely used topology in DC. We emulate the fat-tree topology as listed in Fig.~\ref{figure_implement_NetworkTopo}(b).
	Spine-leaf is another important topology in DC, in which the spine switch (namely core switch) is connected to the leaf switch (namely ToR switch) by the IP network.
	We deploy a 14-node physical cluster that contains 3 switches, on which we emulate a 120-node virtual cluster using the above three topologies and generate the micro and component-level traces.
	The detailed environment is illustrated in Section~\ref{wc-dc}.

Table~\ref{table_all_micro_trace_info} and Table~\ref{table_all_component_trace_info} introduce the micro and component-level traces, respectively, from the perspectives of the corresponding application, pattern, network topology, and switch types. Note that the switch types include ToR, aggregation (agg for short), and core types for three-tier and fat-tree topology, while spine-leaf only includes ToR and core types.
We compute the throughput every fifty milliseconds and mainly classify the network patterns into three categories, including stable, burst, and increase patterns. The stable pattern indicates the throughput is stable within a period of time. The burst pattern means the throughput rises or drops sharply in a short time. The increase pattern means the throughput is gradually rising within a period of time.



\begin{table}[htb]
	\centering
	\caption{Micro Traces in DCNetBench.}
	\label{table_all_micro_trace_info}
	\footnotesize
		\begin{tabular}{cccccc}
			\toprule
			\multicolumn{1}{l}{\textbf{Type}} & \textbf{Traces} & \textbf{Application} & \textbf{Pattern} & \textbf{Topology} & \textbf{Switch} \\
			\midrule
			\multirow{12}{*}{Micro}           & Net\_M1           & Hadoop Grep            & Burst            & Three-tier                    & ToR                  \\
			& Net\_M2           & Hadoop PageRank        & Burst            & Spine-leaf                    & ToR                  \\
			& Net\_M3           & Hadoop PageRank        & Burst            & Spine-leaf                    & ToR                  \\
			& Net\_M4           & Hadoop PageRank        & Burst            & Three-tier                    & ToR                  \\
			& Net\_M5           & Spark RandSample       & Increase         & Spine-leaf                    & Core                 \\
			& Net\_M6           & Image   Classification & Increase         & Three-tier                    & ToR                  \\
			& Net\_M7           & Image   Classification & Increase         & Fat-tree                      & Core                 \\
			& Net\_M8           & Hadoop RandSample      & Increase         & Spine-leaf                    & Core                 \\
			& Net\_M9           & Hadoop Grep            & Stable           & Spine-leaf                    & ToR                  \\
			& Net\_M10          & Spark WordCount        & Stable           & Spine-leaf                    & Core                 \\
			& Net\_M11          & Spark WordCount        & Stable           & Spine-leaf                    & ToR                  \\
			& Net\_M12          & Image   Classification & Stable           & Spine-leaf                    & ToR                   \\    
			\bottomrule
		\end{tabular}
\end{table}
\begin{table}[htb]
	\centering
\caption{Component Traces in DCNetBench.}
\label{table_all_component_trace_info}
		\footnotesize
		\begin{tabular}{cccccc}
			\toprule
			\multicolumn{1}{l}{\textbf{Type}} & \textbf{Traces} & \textbf{Application} & \textbf{Pattern} & \textbf{Topology} & \textbf{Switch} \\
			\midrule
			\multirow{4}{*}{Component}        & Net\_C1           & Image Classification   & Stable, Burst, Increase & Spine-leaf               & ToR                    \\
			& Net\_C2           & Hadoop PageRank        & Increase, Stable       & Spine-leaf               & Core                   \\
			& Net\_C3           & Image Classification   & Increase, Burst         & Fat-tree                 & Core                   \\
			& Net\_C4           & Hadoop WordCount       & Stable, Burst          & Three-tier               & Agg                    \\    
			\bottomrule
		\end{tabular}
\end{table}

\section{Evaluation}

This section evaluates the effectiveness of DCNetBench. Subsection~\ref{eva-setup} introduces the experimental setup. Subsection~\ref{eva-ver} verifies the reality of emulated network traffic. Subsection~\ref{workcharac} performs scaleability characterization of DCNetBench. Subsection~\ref{switch-eva} evaluates the switch chips using micro traces.

\subsection{Experimental Setup}~\label{eva-setup}
We deploy the experimental environments for reality verification, scaleability characterization, and switch chip evaluation. They use the same software versions: Ubuntu 20.04 LTS, Python 3.8.10, Docker 20.10.13, MaxiNet version 1.2~\footnote{We built the emulated system based on this version.}, Open VSwitch 2.13, TCPDUMP 4.9.3, Tcpreplay 4.3.2, Hadoop 2.7.1, Spark 1.6.0. The cluster information are as follows.
\subsubsection{Setup for Reality Verification} ~\label{verify_setup}
We deploy an 8-node physical cluster and an 8-node emulated cluster to verify their similarity.
Each physical node contains two CPU processors, 
each of which has six physical cores, 512 KB L1 cache, and 2 MB L2 cache. We enable Hyper-threading. 
The memory is 32 GB. All the nodes are connected to a 1 Gb Ethernet network.
The emulated cluster is deployed on the 8-node physical cluster. 

\subsubsection{Setup for Scaleability Characterization}~\label{wc-dc}
\begin{figure}[htb]
	\centering
	\includegraphics[width=0.6\linewidth]{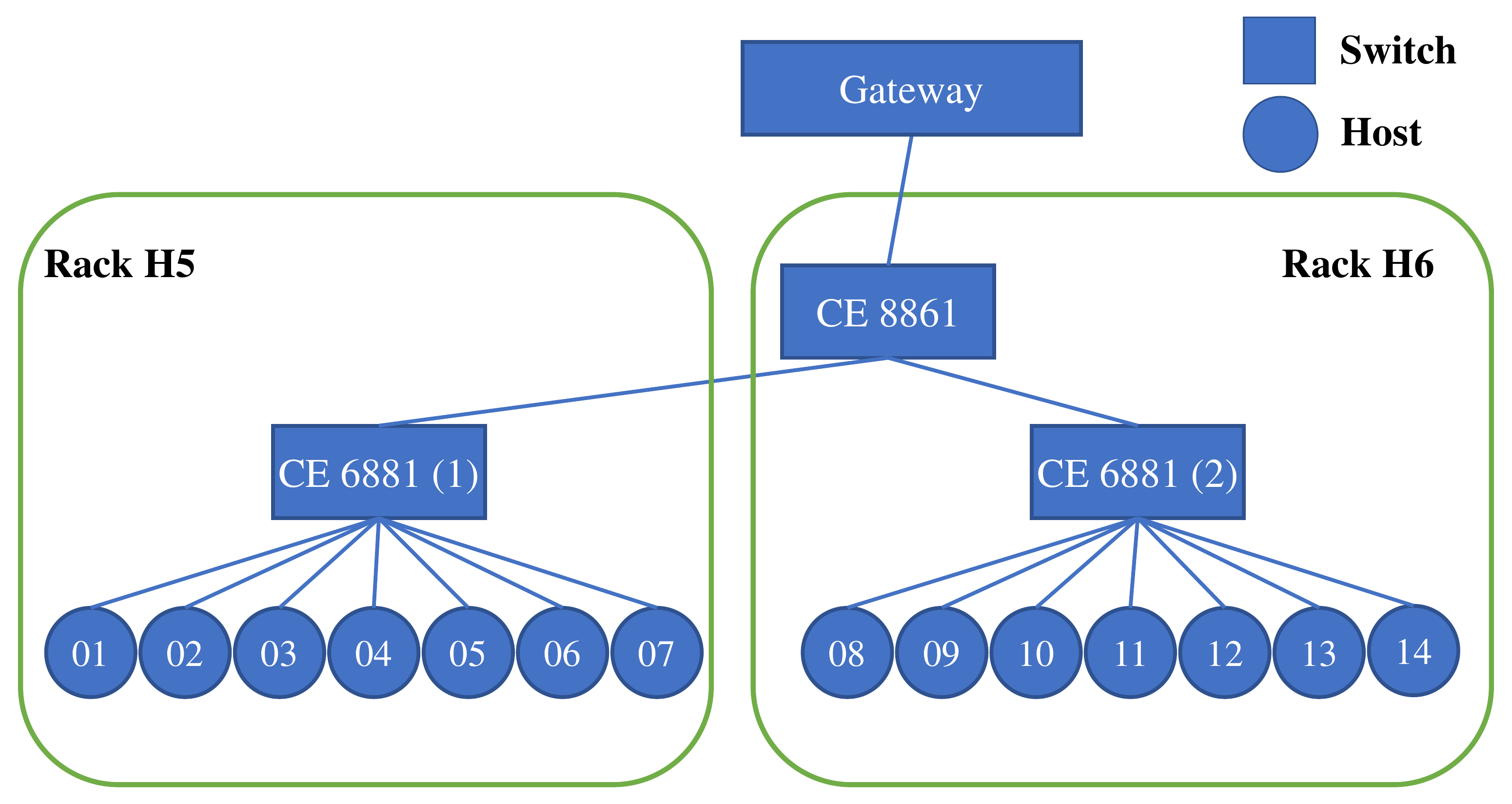}
	\caption{Topology of 14-node Physical Cluster.}
	\label{figure_generation_physical_network}
\end{figure}

We characterize DCNetBench using the automatic framework to generate traces according to the different user-defined inputs. We evaluate its scaleability by using six various applications and three important network topologies. The experimental environment is on a 120-node virtual cluster that is emulated on a 14-node physical cluster. 
Each physical node is equipped with a CPU processor with 12 physical cores, a base frequency of 2.1 GHz, and a max turbo frequency of 3 GHz. 
The memory and disk storages are 32 GB and 2.2 TB, respectively.
The physical nodes are connected through three switches:  a Huawei CE8861 and two CE 6881 devices.
The topology of the physical cluster is shown in Fig. ~\ref{figure_generation_physical_network}.
We use the optical module to connect servers and switches, and set all ports' bandwidth as 10 Gbs.
The emulated 120-node cluster settings are shown in Fig.~\ref{figure_implement_NetworkTopo}.

\subsubsection{Setup for Switch Chip Evaluation}
We evaluate five typical switches using micro traces on a 13-node cluster.
The server information is the same as the servers in~\ref{verify_setup}.
The switch specifications are shown in Table ~\ref{table_evaluation_device_info}.
For each switch evaluation, we use Cat5E twisted pair to connect all servers to this switch with a star topology. 

\begin{table}[htb]
	\centering
	\caption{The Overview of Five Switch Chips.}
	\label{table_evaluation_device_info}
	\resizebox{1\columnwidth}{!} {
	\begin{tabular}{cccccc}
		\toprule
		\textbf{Series}         & \textbf{Device model}       & \textbf{Port type}            & \textbf{Exchange capacity} & \textbf{Sending speed}  \\
		\midrule
		HUAWEI S7706   & Quidway S7706      & 10/100/1000 Mbps & 76.8 Tbps           & 8640 Mpps      \\
		HUAWEI S5710   & S5710-52C-EI       & 10/100/1000 Mbps & 416 Gbps            & 192 Mpps       \\
		H3C S5120      & S5120-48P-EI       & 10/100/1000 Mbps & 240 Gbps            & 72 Mpps        \\
		HUAWEI S5324TP & Quidway S5324TP-SI & 10/100/1000 Mbps & 48 Gbps             & 36 Mpps        \\
		CISCO SRW2024 & SRW2024 & 10/100/1000 Mbps & 48 Gbps             & 36 Mpps        \\
		\bottomrule
	\end{tabular}
	}
\end{table}
		
\subsection{The Reality of Emulated Traffic}~\label{eva-ver}
We conduct experiments to verify that the emulated environment has little impact on traffic patterns compared to a physical one. 
We use three representative big data applications, i.e., Grep, PageRank, and MatrixMult, implemented with Hadoop and Spark. They have different computation complexities and process different data like text, graphs, and matrices. We run these workloads on the physical and emulated cluster illustrated in Section ~\ref{verify_setup} and monitor the network traffic using the TCPDUMP tool. We use two significant metrics -- cumulative distribution of flow (CDF) and traffic matrix -- to verify the reality from the perspectives of times series similarity and spatial similarity, respectively. 




\begin{figure}[htbp]
	\subfloat[Hadoop Grep]{
		\includegraphics[width = 0.29\linewidth]{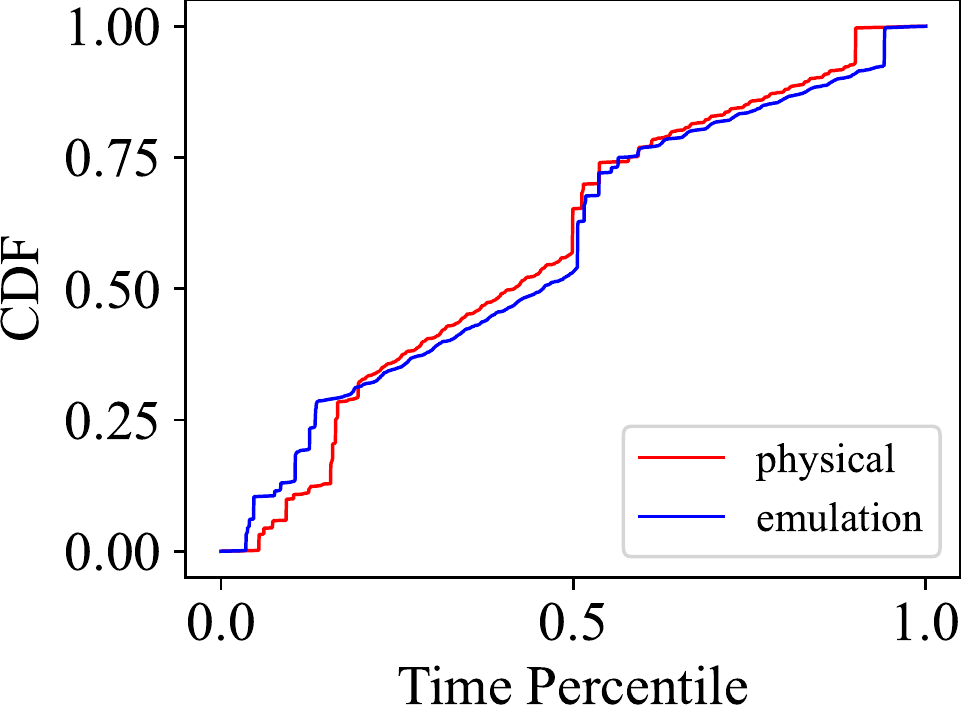}
		\label{figure_validity_CDF_h_grep_2GB_1}
	} \quad
	\subfloat[Hadoop MatrixMult]{
		\includegraphics[width = 0.29\linewidth]{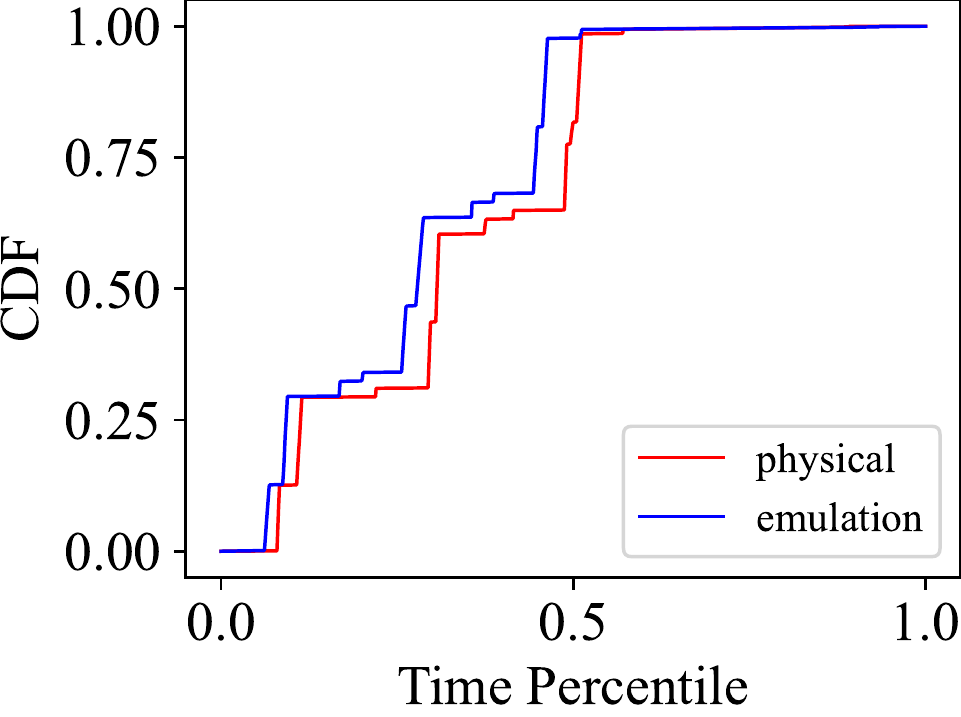}
		\label{figure_validity_CDF_h_MatrixMult_4k_1}
	} \quad	
	\subfloat[Hadoop PageRank] {
		\includegraphics[width = 0.29\linewidth]{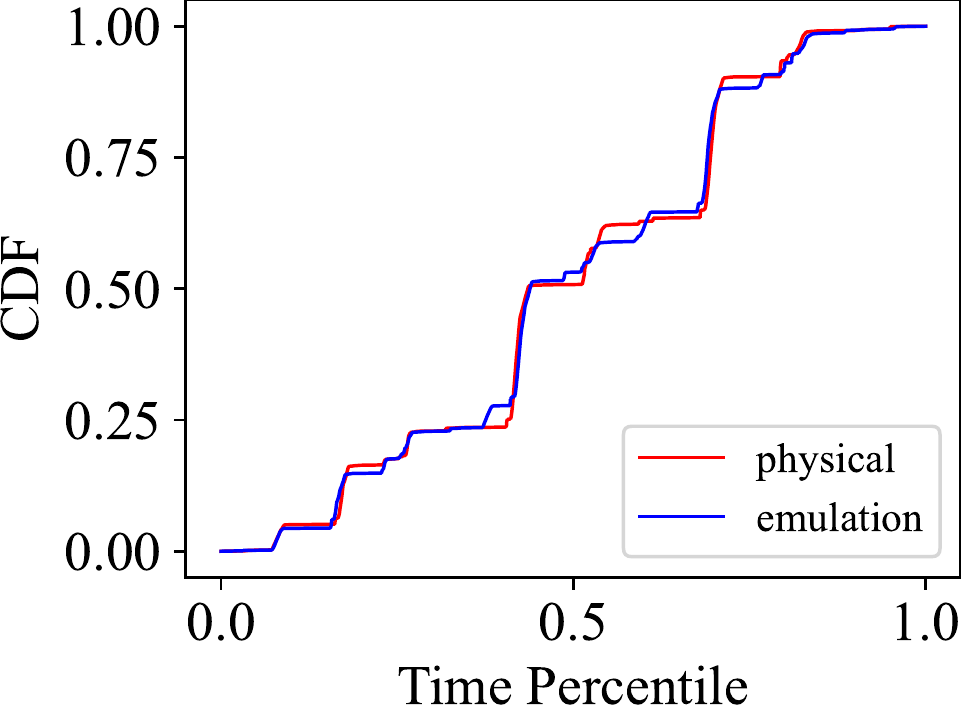}
		\label{figure_validity_CDF_h_PageRank_20_3}
	} \\

	\subfloat[Spark Grep] {
		\includegraphics[width = 0.29\linewidth]{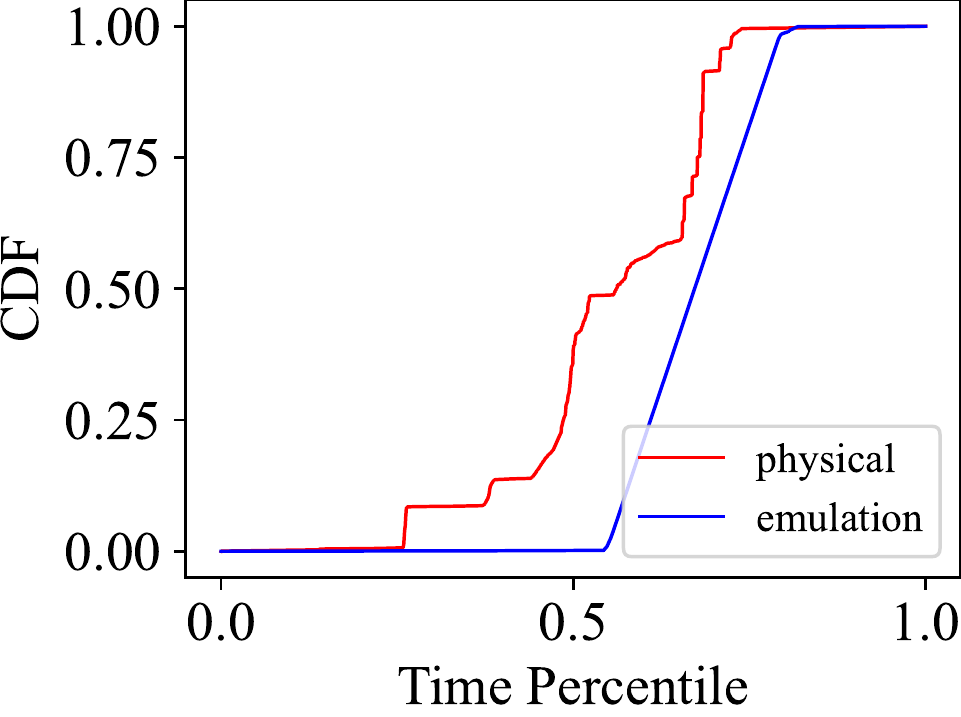}
		\label{figure_validity_CDF_s_grep_2GB_1}
	} \quad
	\subfloat[Spark MatrixMult]{
		\includegraphics[width = 0.29\linewidth]{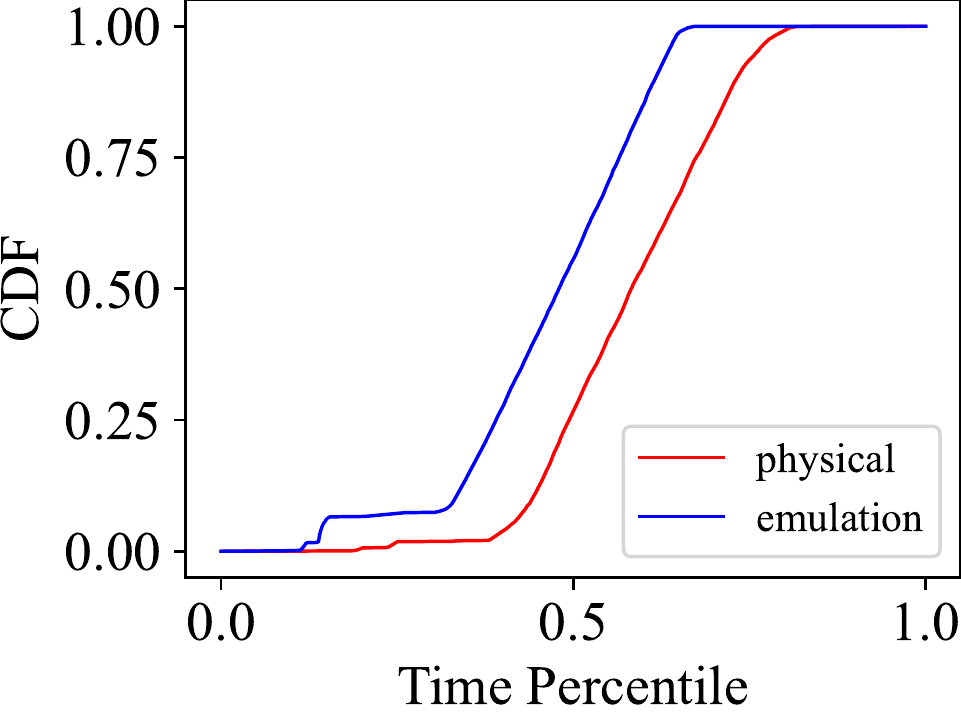}
		\label{figure_validity_CDF_s_MatrixMult_4k_1}
	} \quad
	\subfloat[Spark PageRank]{
		\includegraphics[width = 0.29\linewidth]{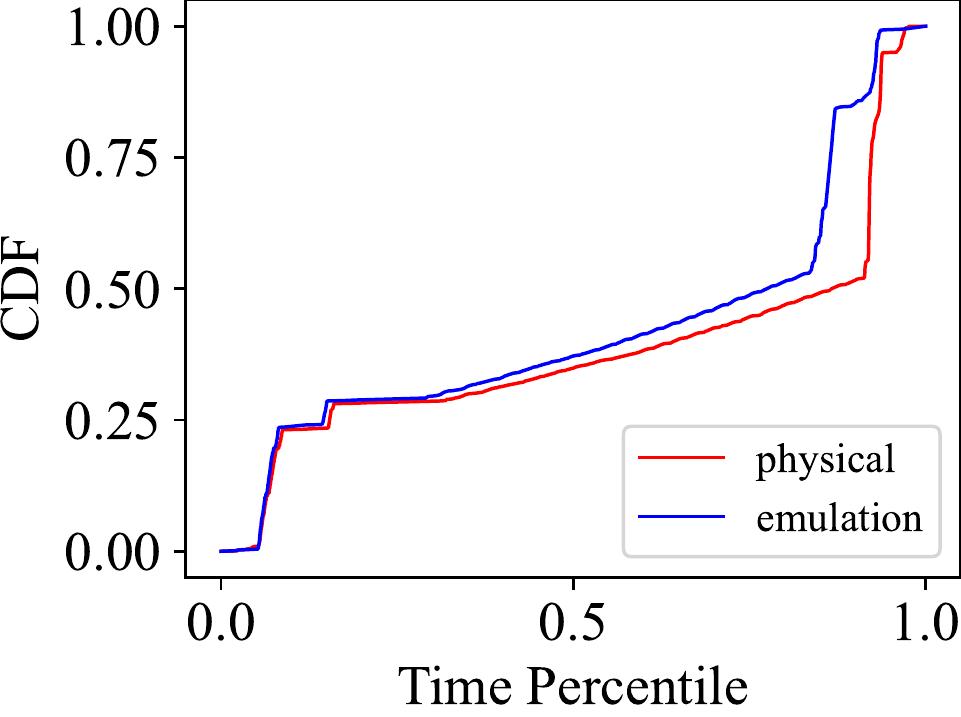}
		\label{figure_validity_CDF_s_PageRank_20_1}
	}
	\caption{Cumulative Distribution of Flow (CDF) in Emulated and Physical Environments.}
	\label{figure_validity_CDF}
\end{figure}


Fig. ~\ref{figure_validity_CDF} shows the time-series CDF similarity of the physical and emulated network traffic.
Since they have different execution time, we normalize the current time using its total running time as a baseline. The value ranges from 0 to 1, which means the percentage that occupies the total time. 
We find that they have consistent CDF trends, indicating the physical and emulated traffic have similar time-series patterns.


\begin{figure}[htb]
	\centering
	\subfloat[Hadoop MatrixMult]{
		\includegraphics[width = 0.22\linewidth]{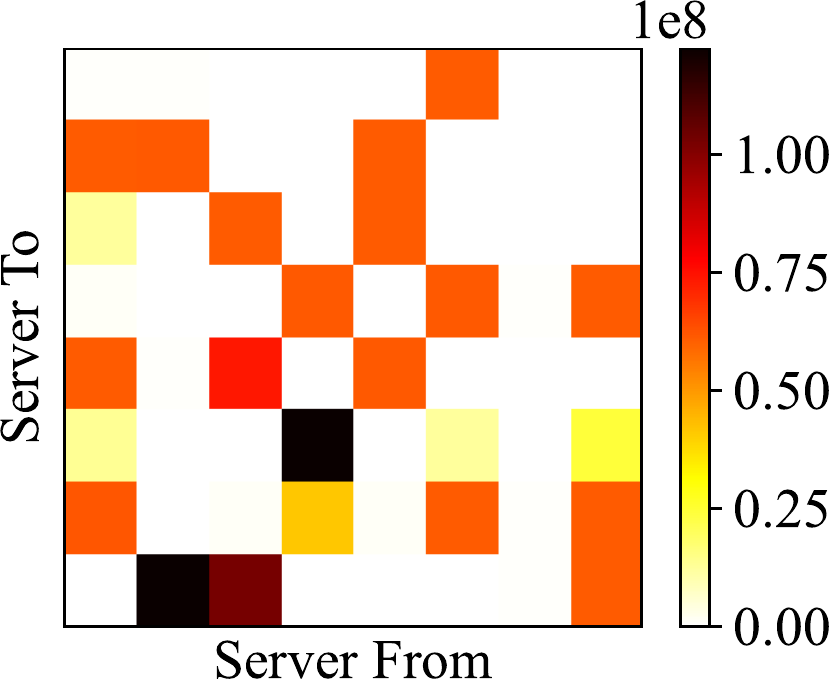} \quad
		\includegraphics[width = 0.22\linewidth]{figure_traffic_matrix_h_MatrixMult_4k_1_physical.pdf}
		\label{figure_validity_CDF_h_grep_2GB_1}
	} \quad
	\subfloat[Spark PageRank]{
		\includegraphics[width = 0.22\linewidth]{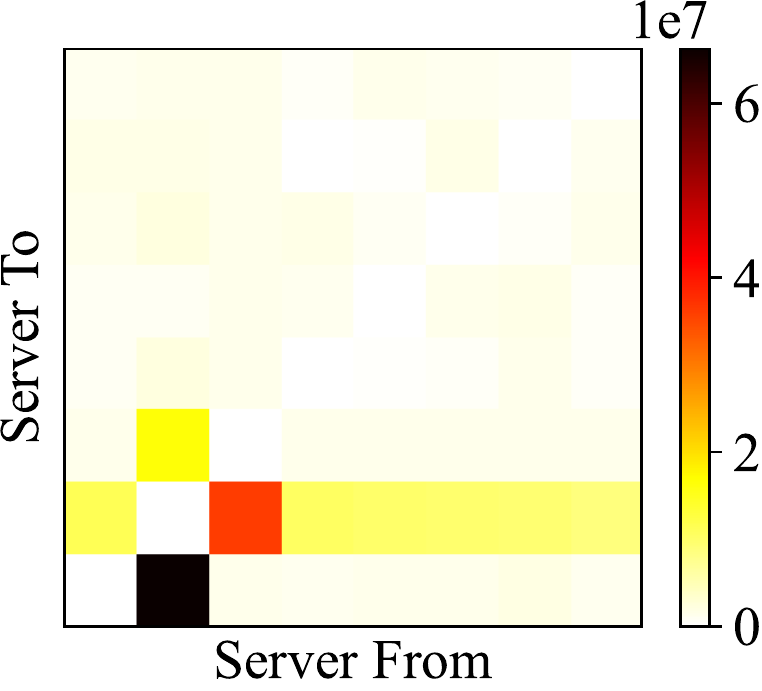} \quad
		\includegraphics[width = 0.22\linewidth]{figure_traffic_matrix_s_PageRank_20_1_physical.pdf}
		\label{figure_traffic_matrix_s_PageRank_20_1}
	} \\
	\caption{Traffic Matrix in Emulated (left) and Physical (right) Environments.}
	\label{figure_validity_TrafficMatric}
\end{figure}


Fig. ~\ref{figure_validity_TrafficMatric} shows the traffic matrix similarity of the physical and emulated network traffic. Note that due to the page limitation, we only present two applications -- Hadoop MatrixMult and Spark PageRank -- with high complexity and different input data. The traffic matrix reflects the network transmission from a server node to another node and a darker color means a larger packet transmission. We find that they share similar traffic matrix patterns. Note that their shapes may have slight differences due to the permutations of the nodes on the coordinate axis are not totally the same. For example, the transmission of node A to node B is represented by a coordinate in the traffic matrix. However, the coordinate in the physical matrix and emulated matrix is different. 

Above all, we conclude that the emulated network traffic is similar to the physical one from both the time series and spatial dimensions. The emulation approach is practical to capture the network traffic patterns of applications.

\subsection{Scaleability Characterization on DCNetBench}~\label{workcharac}

We characterize the scaleability of DCNetBench by using six various applications: Grep, WordCount, Random Sampling, PageRank,  Matrix Multiplication, and Image Classification shown in Table~\ref{implement_all_workloads_info} and three important network topologies: three-tier tree, fat-tree, and spine-leaf shown in Fig.~\ref{figure_implement_NetworkTopo}.
We generate the micro and component-level traces using DCNetBench as shown in Table~\ref{table_all_micro_trace_info}. 
The patterns of different traces show that different applications and DC environments impact the network traffic. Thus, scaleability is essential for DC network benchmarking and DCNetBench can achieve scaleability.

\begin{figure*}[!htb]
	\subfloat[Net\_M1]{
		\includegraphics[width = 0.13\linewidth]{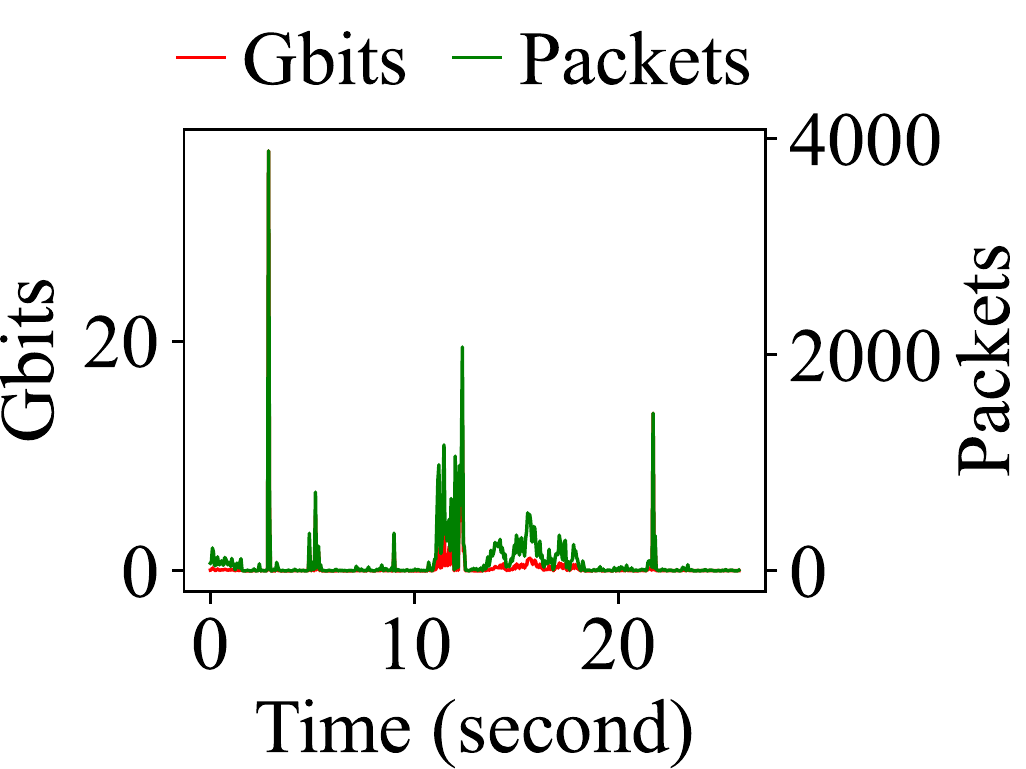}
		\label{figure_testcase_net_M1}
	} \quad
	\subfloat[Net\_M2]{
		\includegraphics[width = 0.13\linewidth]{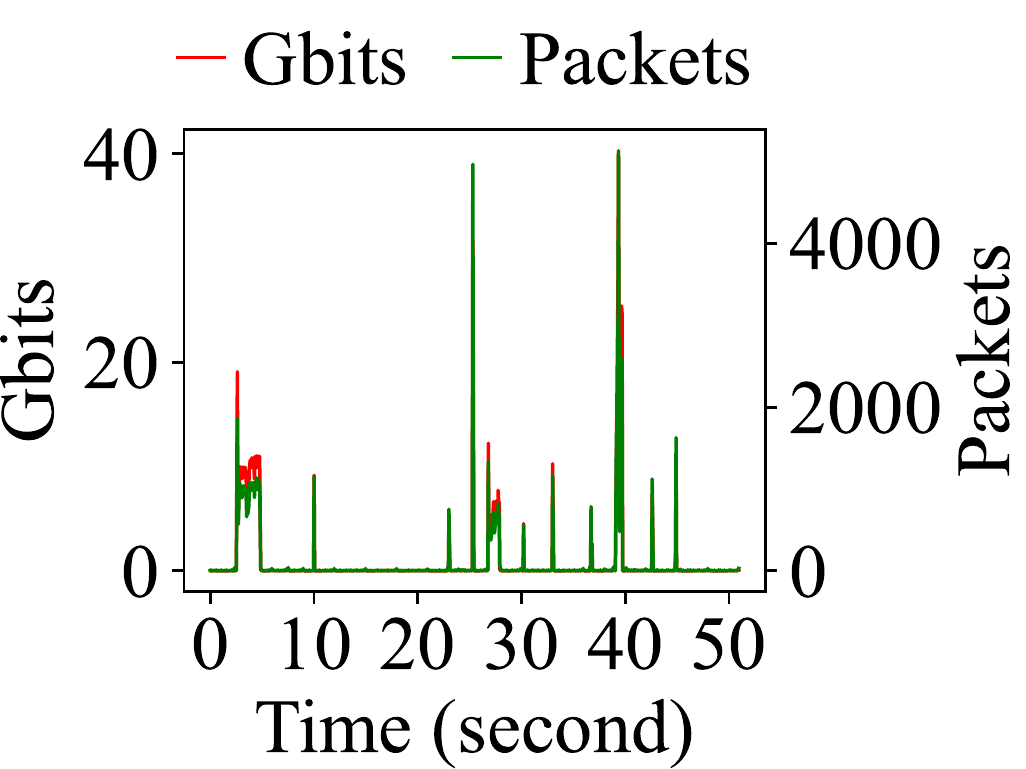}
		\label{figure_testcase_net_M2}
	} \quad        
	\subfloat[Net\_M3] {
		\includegraphics[width = 0.13\linewidth]{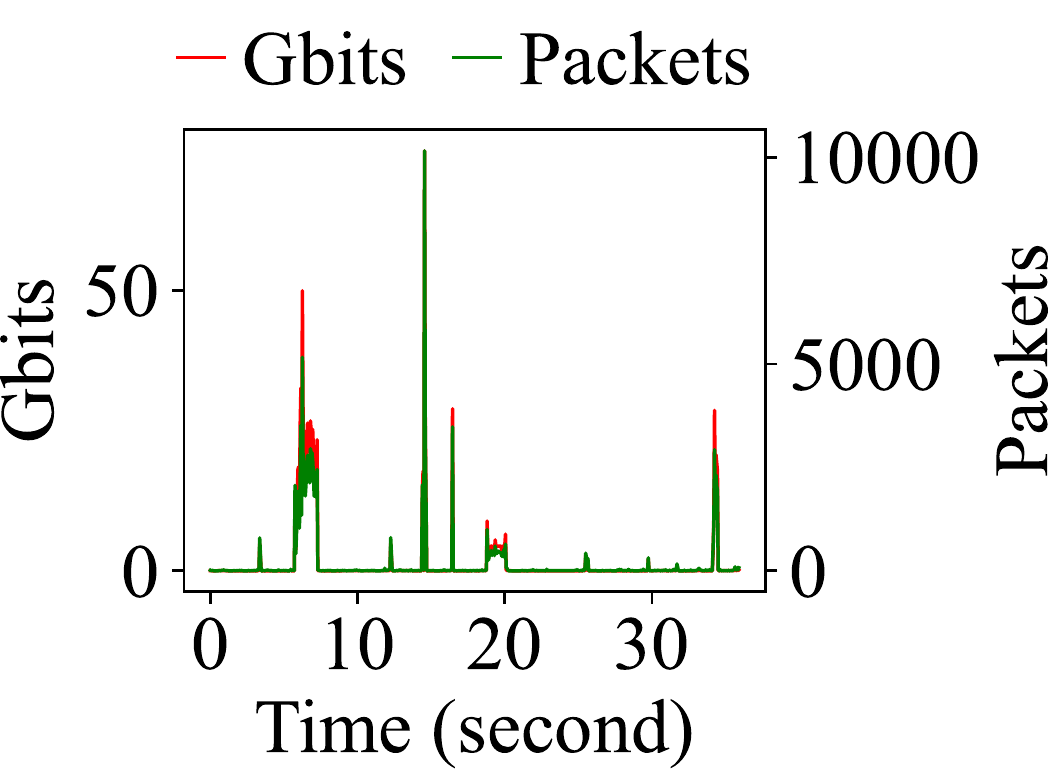}
		\label{figure_testcase_net_M3}
	} \quad
	\subfloat[Net\_M4] {
		\includegraphics[width = 0.13\linewidth]{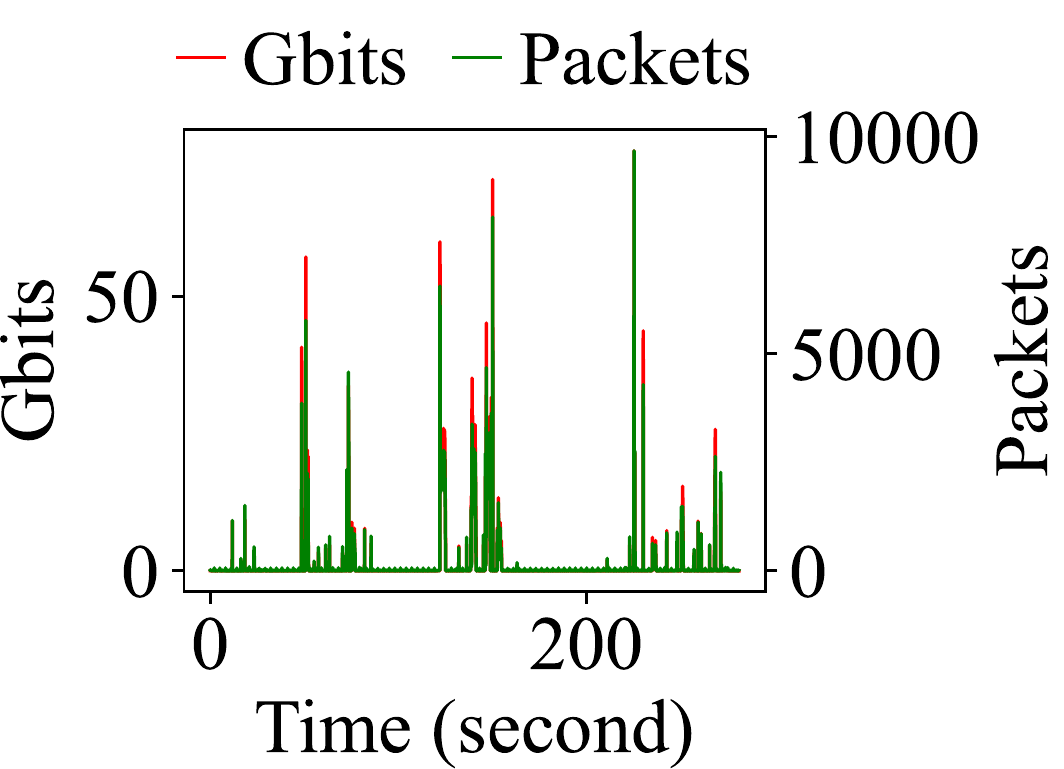}
		\label{figure_testcase_net_M4}
	} \quad 
	\subfloat[Net\_M5]{
		\includegraphics[width = 0.13\linewidth]{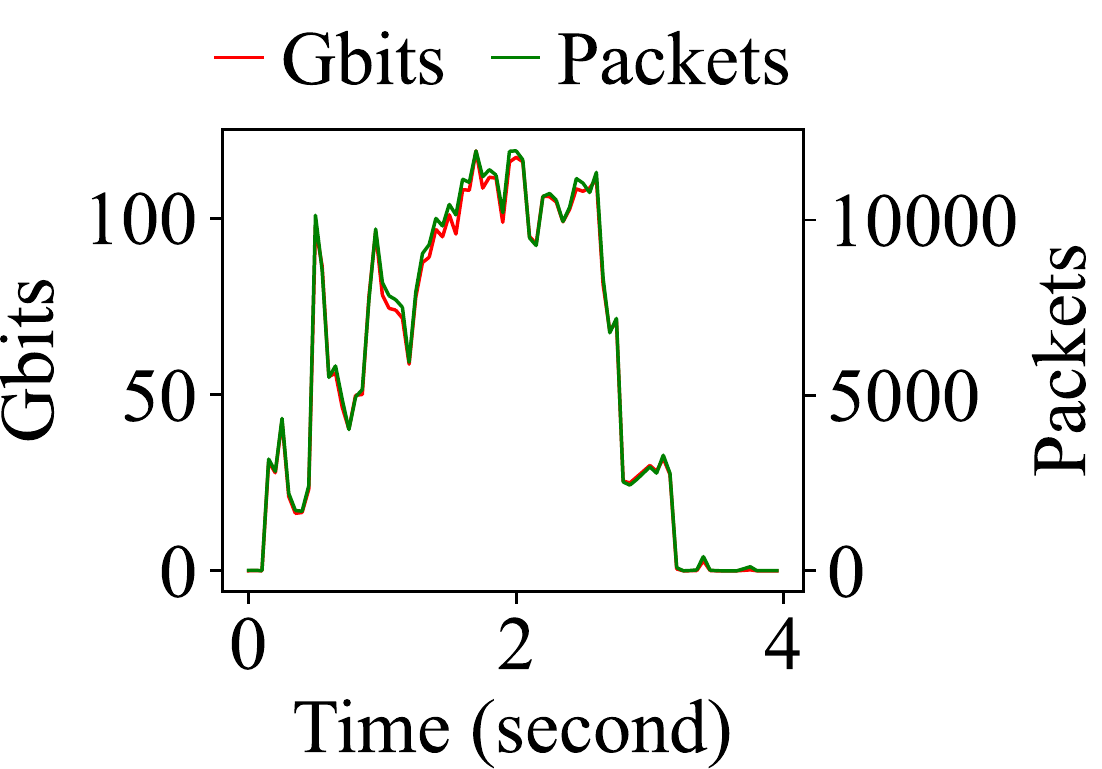}
		\label{figure_testcase_net_M5}
	} \quad
	\subfloat[Net\_M6]{
		\includegraphics[width = 0.13\linewidth]{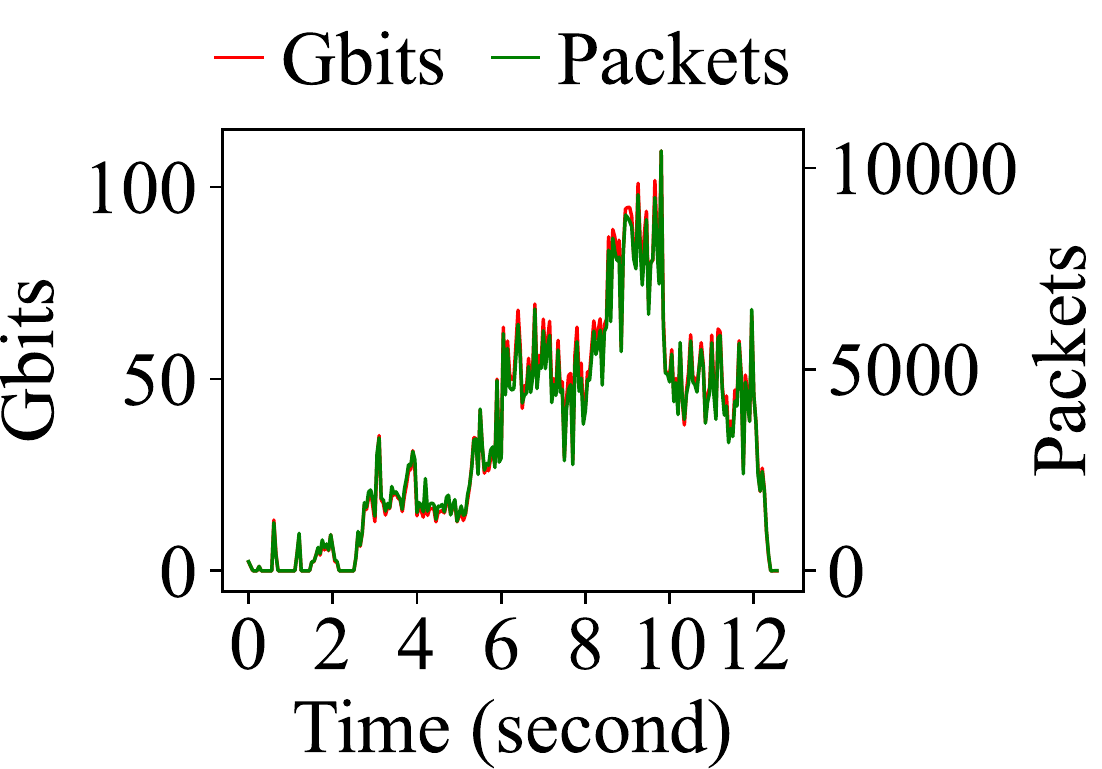}
		\label{figure_testcase_net_M6}
	} \\
	
	\subfloat[Net\_M7] {
		\includegraphics[width = 0.13\linewidth]{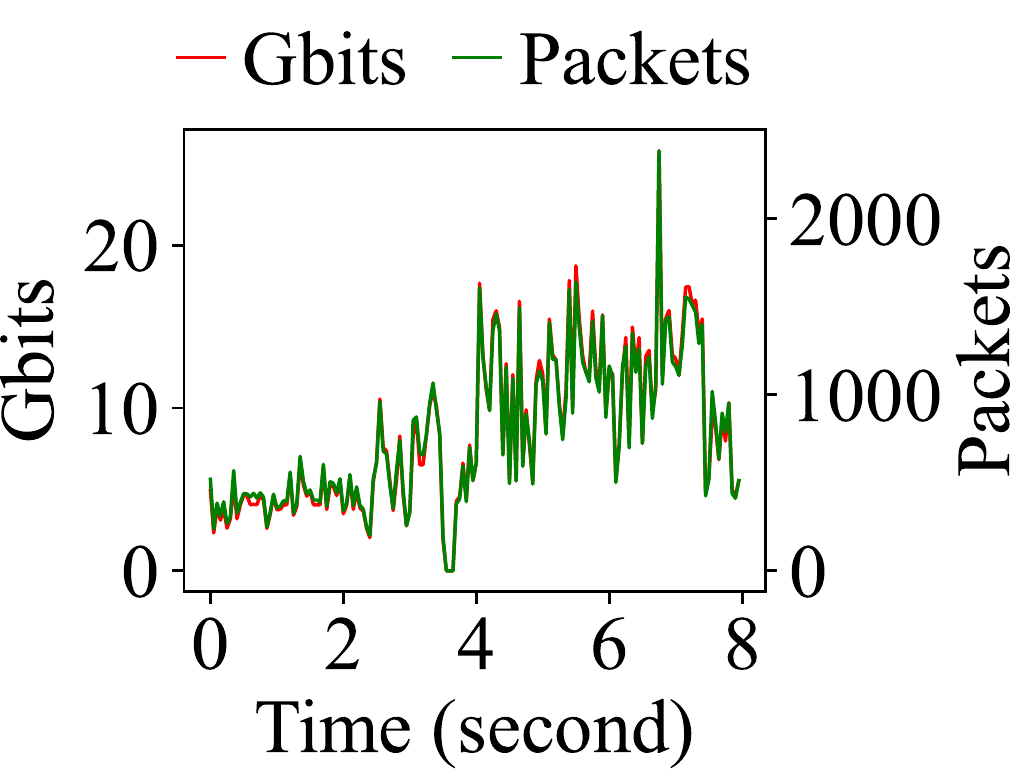}
		\label{figure_testcase_net_M7}
	} \quad
	\subfloat[Net\_M8] {
		\includegraphics[width = 0.13\linewidth]{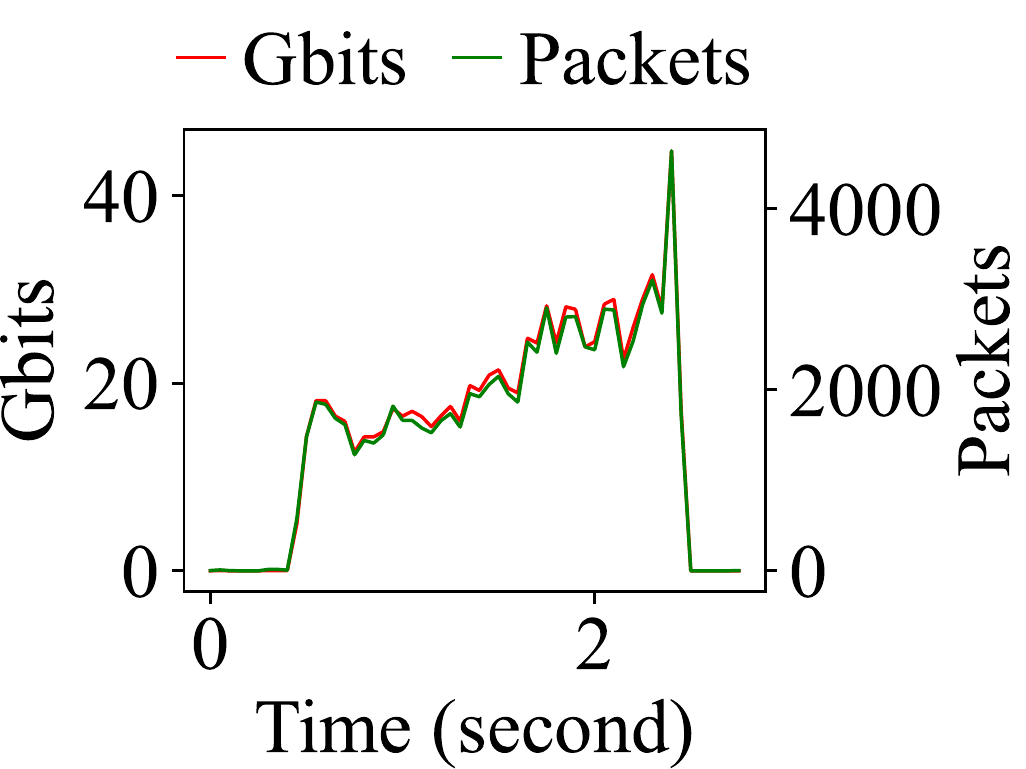}
		\label{figure_testcase_net_M8}
	} \quad
	\subfloat[Net\_M9]{
		\includegraphics[width = 0.13\linewidth]{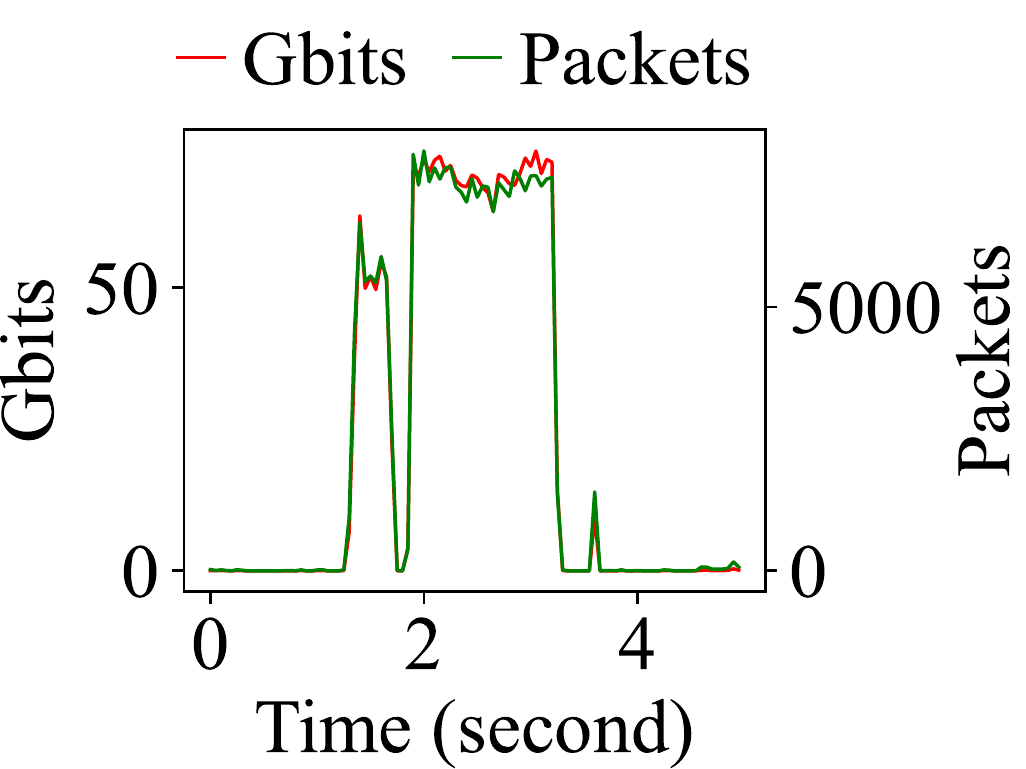}
		\label{figure_testcase_net_M9}
	} \quad
	\subfloat[Net\_M10]{
		\includegraphics[width = 0.13\linewidth]{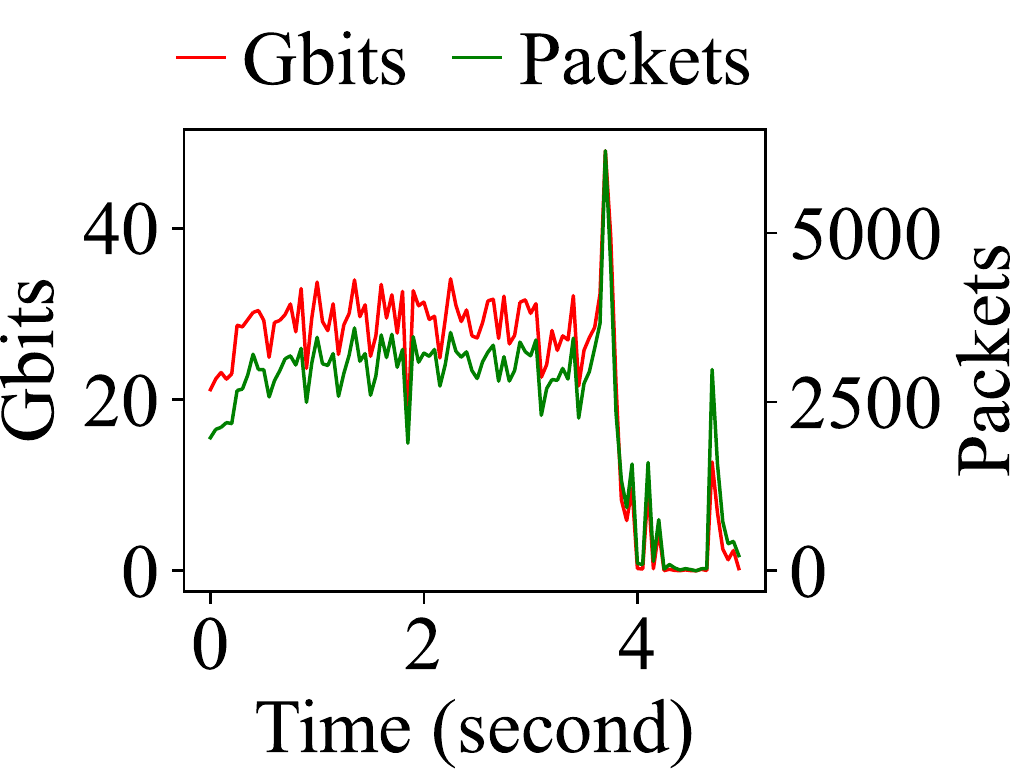}
		\label{figure_testcase_net_10}
	} \quad        
	\subfloat[Net\_M11] {
		\includegraphics[width = 0.13\linewidth]{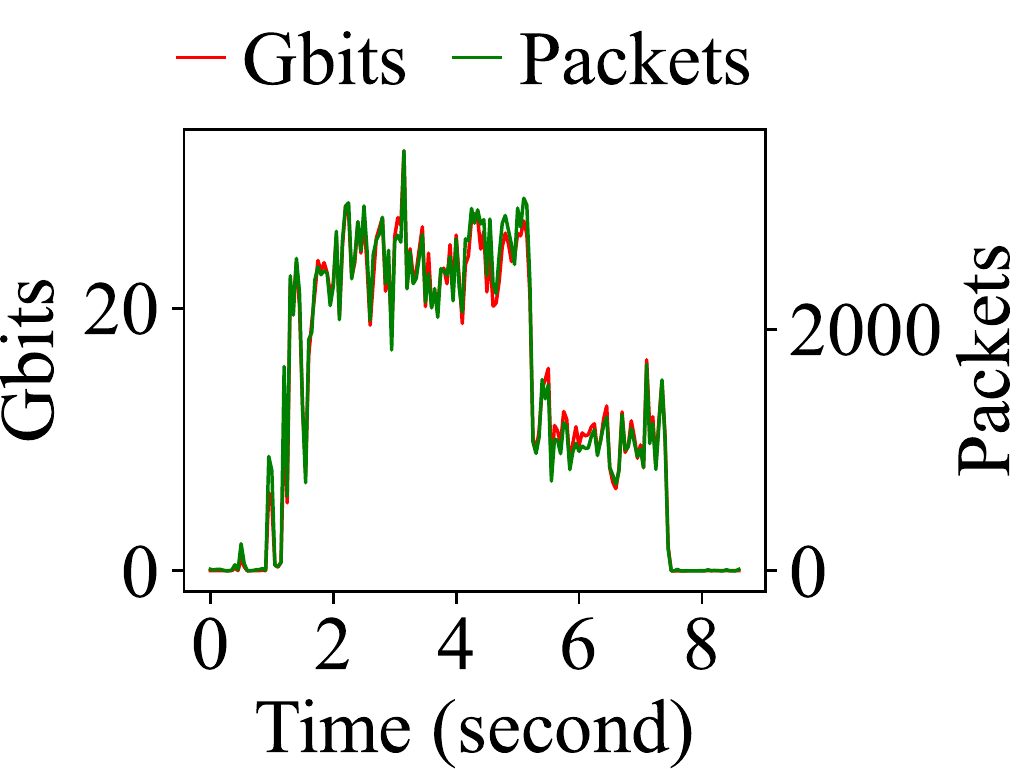}
		\label{figure_testcase_net_M11}
	} \quad
	\subfloat[Net\_M12] {
		\includegraphics[width = 0.13\linewidth]{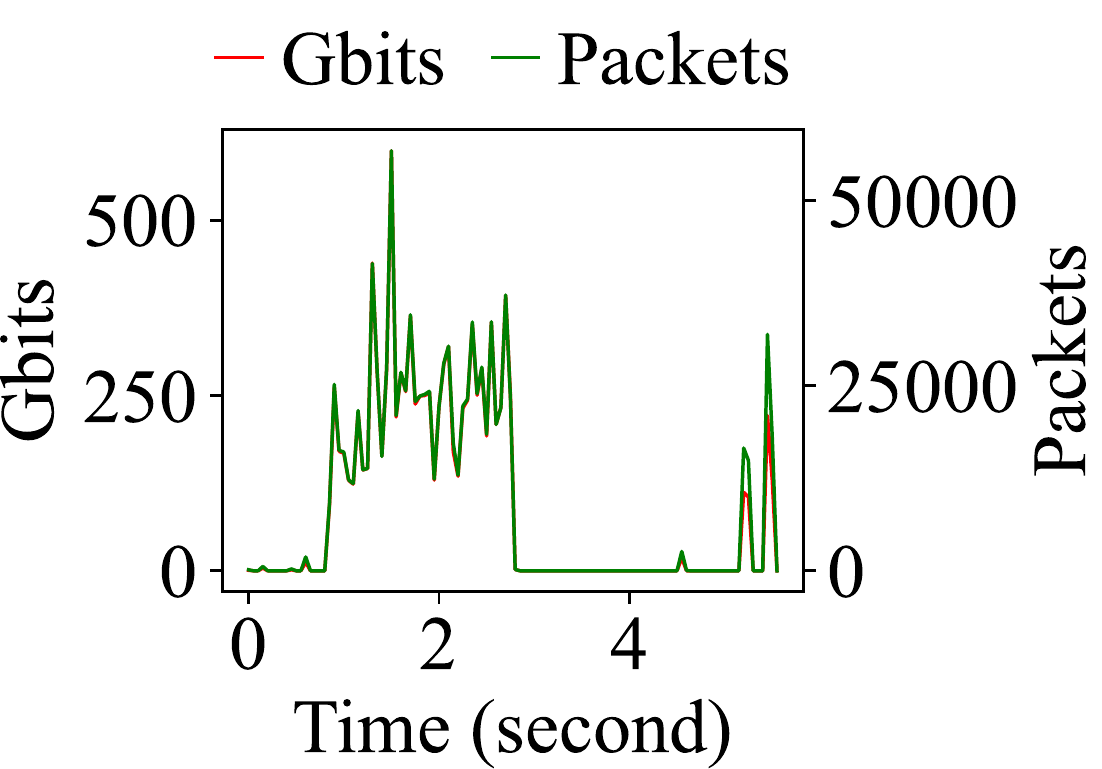}
		\label{figure_testcase_net_M12}
	} \\
	
	\caption{Time-series Patterns of Micro Traces.}
	\label{figure_testcase_CDF_all}
\end{figure*}

We analyze twelve micro traces, each of which represents the most typical network traffic pattern of the corresponding application. We first analyze the time-series patterns of the network traffic. We monitor the data transmission during the whole running time and count the number of packets and the size of packets every 50 milliseconds.
Fig.~\ref{figure_testcase_CDF_all} shows the analyzed result. The left y-axis shows the variation of packet size along with the time (Gbits in Fig.~\ref{figure_testcase_CDF_all}) and the right y-axis shows the variation of the number of packets along with the time (Packets in Fig.~\ref{figure_testcase_CDF_all}). We find that the traces Net\_M1 to Net\_M4 conform to the burst pattern. Meanwhile, they reflect diverse burst frequencies and burst amplitude. For example, the Net\_M1 has the lowest frequency and diminishing burst amplitude compared to the other three. The Net\_M4 has the highest burst frequency and amplitude. 
The traces Net\_M5 to Net\_M8 conform to the increase pattern. Meanwhile, they reflect various slopes and fluctuations. For example, the Net\_M5 has the highest slopes while the Net\_M7 and Net\_M8 increase smoothly. The Net\_M6 and Net\_M7 have more frequent fluctuations.
The traces Net\_M9 to Net\_M12 conform to the stable pattern. Meanwhile, they reflect diverse throughput rates ranging from 400 to 5000 Gbits per second. For example, the Net\_M12 has the highest throughput rates (5000) while the Net\_M10 (400) has the lowest. Besides, the Net\_M11 contains two-stage stable patterns with a higher throughput rate of 600 Gbits per second and a lower value of 200. 

\begin{figure*}[htb]
	\subfloat[Net\_M1]{
		\includegraphics[width = 0.13\linewidth]{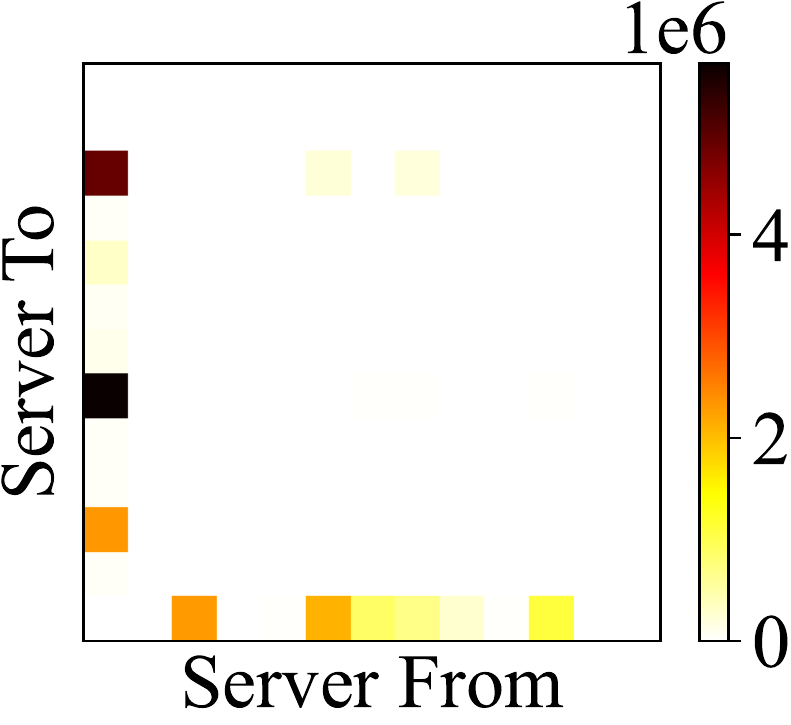}
		\label{figure_testcase_net_M1}
	} \quad
	\subfloat[Net\_M2]{
		\includegraphics[width = 0.13\linewidth]{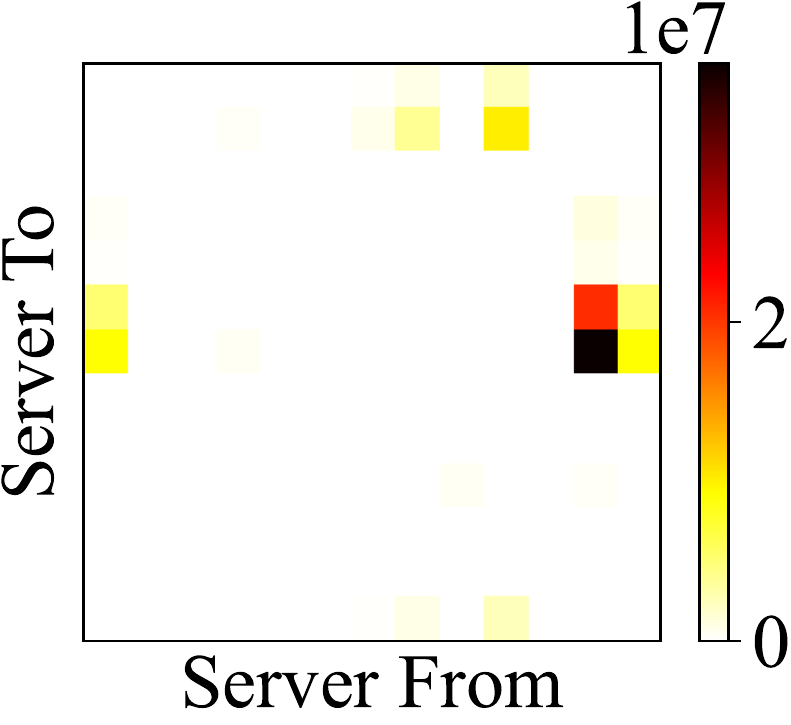}
		\label{figure_testcase_net_M2}
	} \quad        
	\subfloat[Net\_M3] {
		\includegraphics[width = 0.13\linewidth]{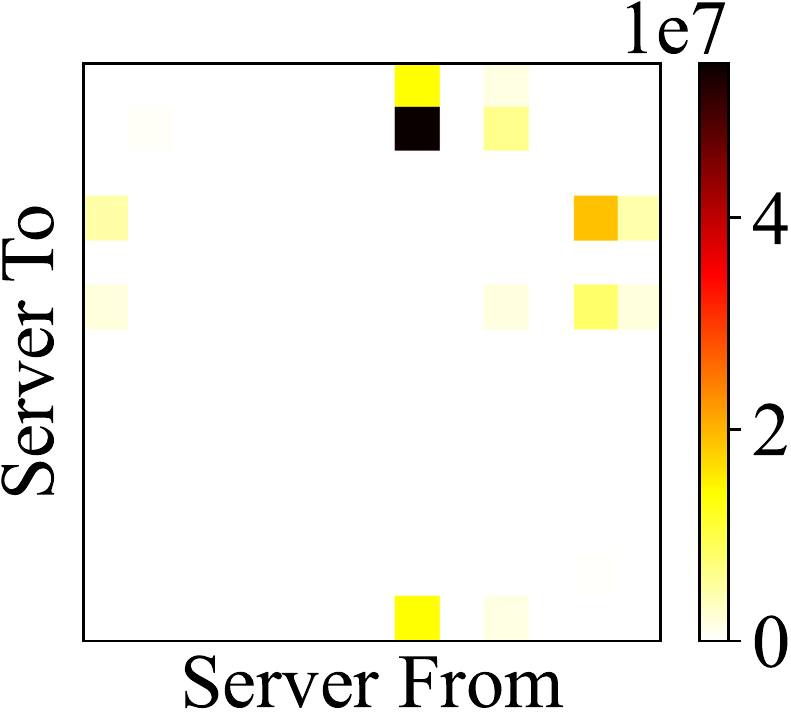}
		\label{figure_testcase_net_M3}
	} \quad
	\subfloat[Net\_M4] {
		\includegraphics[width = 0.13\linewidth]{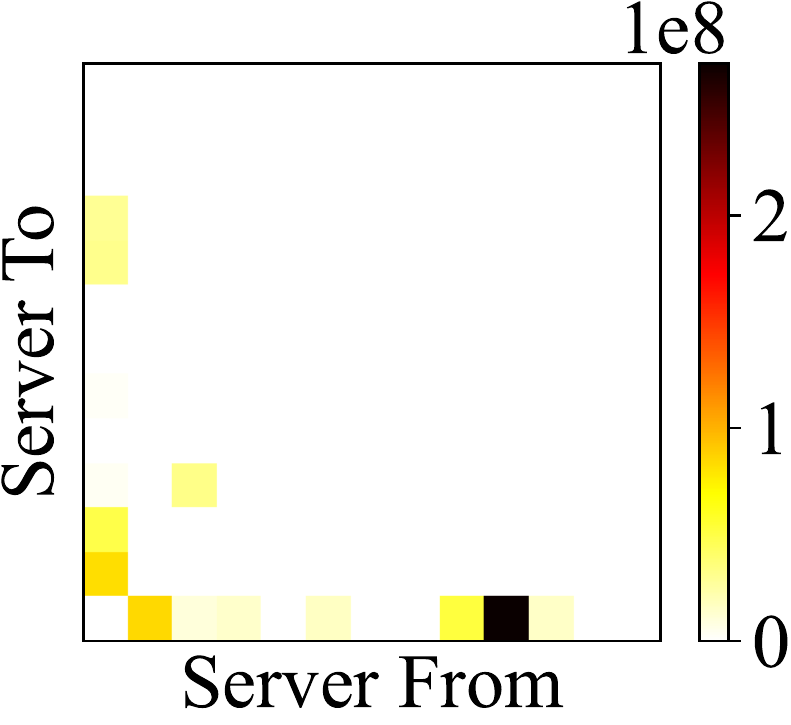}
		\label{figure_testcase_net_M4}
	} \quad 
	\subfloat[Net\_M5]{
		\includegraphics[width = 0.13\linewidth]{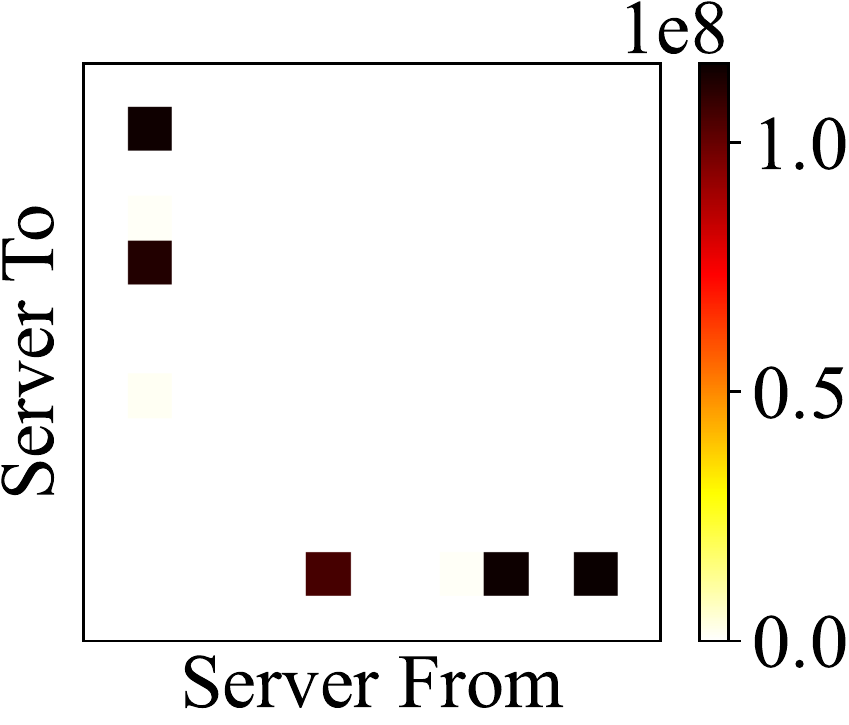}
		\label{figure_testcase_net_M5}
	} \quad
	\subfloat[Net\_M6]{
		\includegraphics[width = 0.13\linewidth]{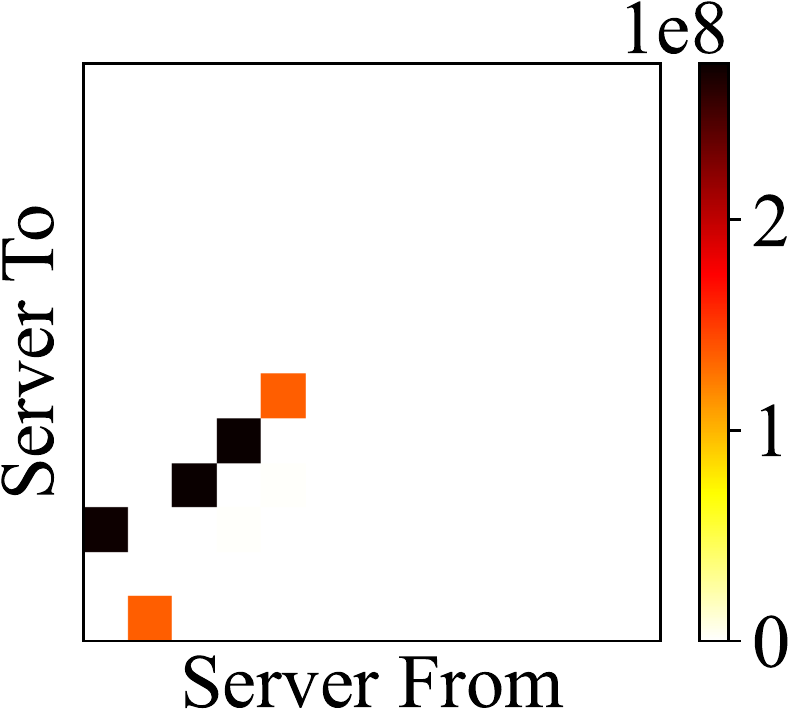}
		\label{figure_testcase_net_M6}
	} \\
	
	\subfloat[Net\_M7] {
		\includegraphics[width = 0.13\linewidth]{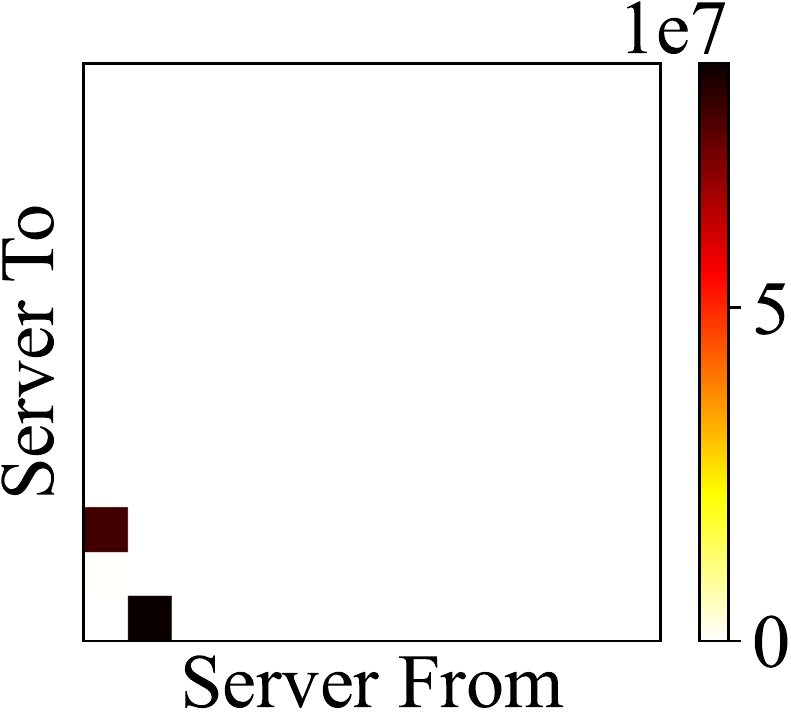}
		\label{figure_testcase_net_M7}
	} \quad
	\subfloat[Net\_M8] {
		\includegraphics[width = 0.13\linewidth]{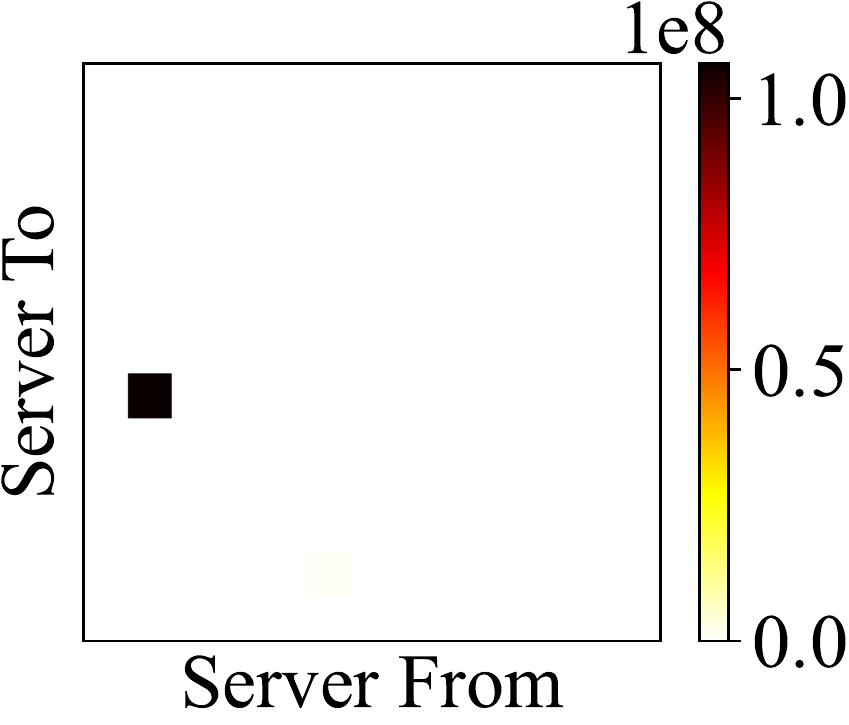}
		\label{figure_testcase_net_M8}
	} \quad
	\subfloat[Net\_M9]{
		\includegraphics[width = 0.13\linewidth]{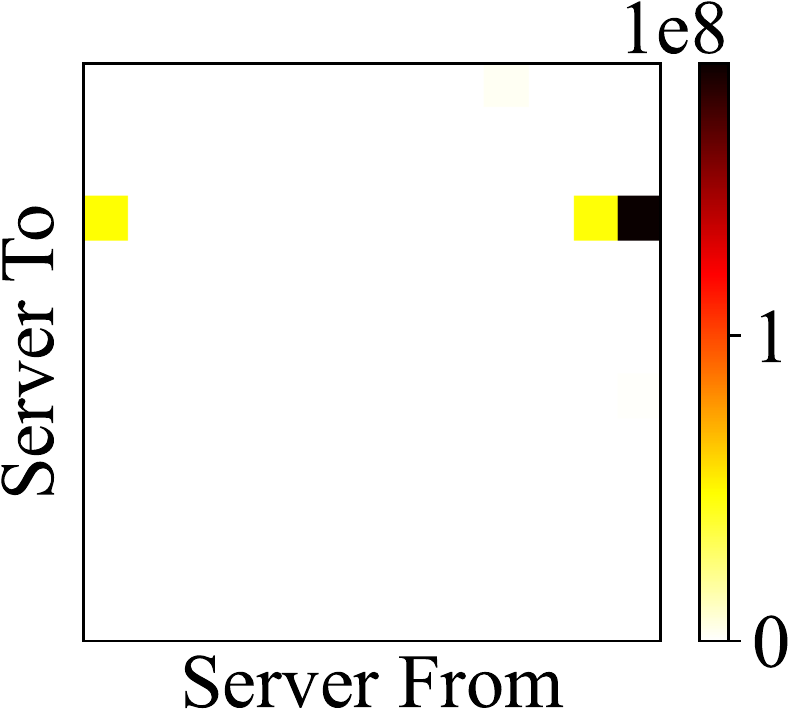}
		\label{figure_testcase_net_M9}
	} \quad
	\subfloat[Net\_M10]{
		\includegraphics[width = 0.13\linewidth]{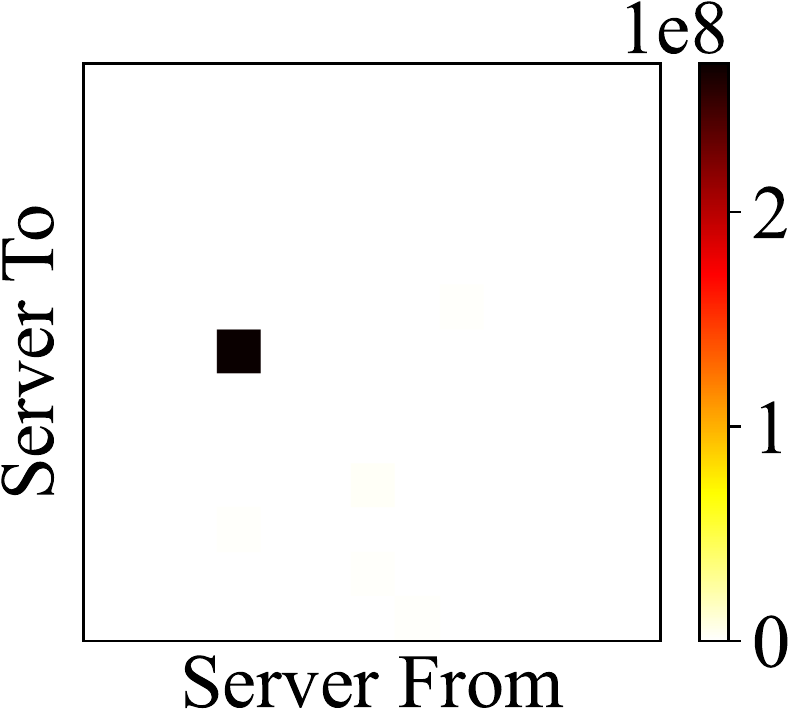}
		\label{figure_testcase_net_10}
	} \quad        
	\subfloat[Net\_M11] {
		\includegraphics[width = 0.13\linewidth]{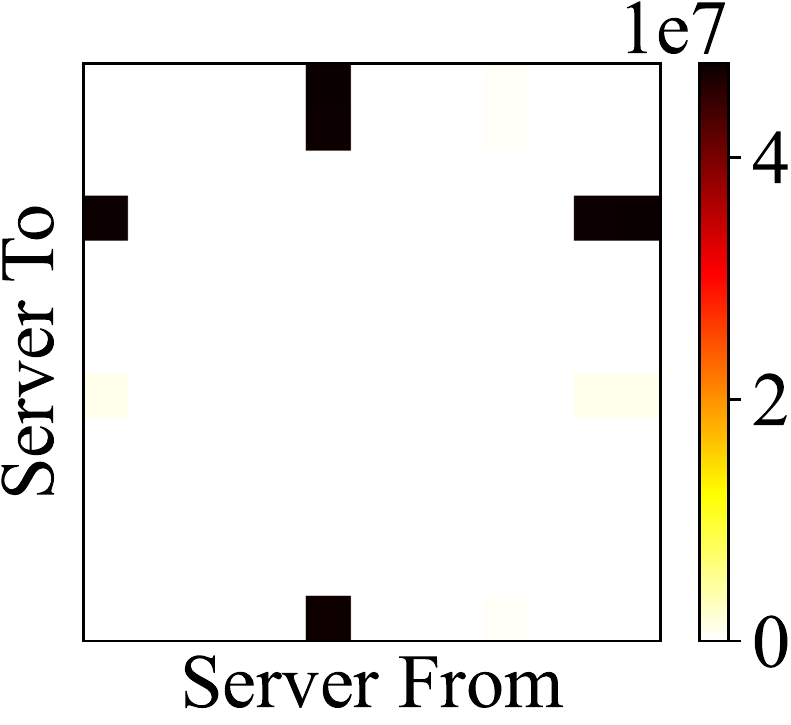}
		\label{figure_testcase_net_M11}
	} \quad
	\subfloat[Net\_M12] {
		\includegraphics[width = 0.13\linewidth]{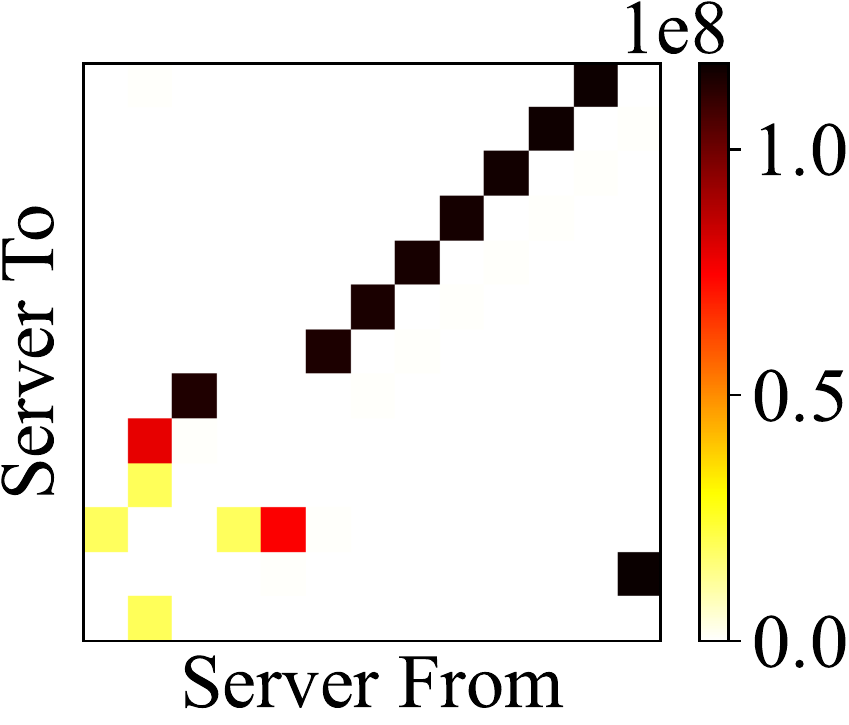}
		\label{figure_testcase_net_M12}
	} \\
	
	\caption{Traffic Matrix Patterns of Micro Traces.}
	\label{figure_testcase_traffic_matrix_all}
\end{figure*}

\begin{figure*}[htb]
	\subfloat[Net\_M1]{
		\includegraphics[width = 0.13\linewidth]{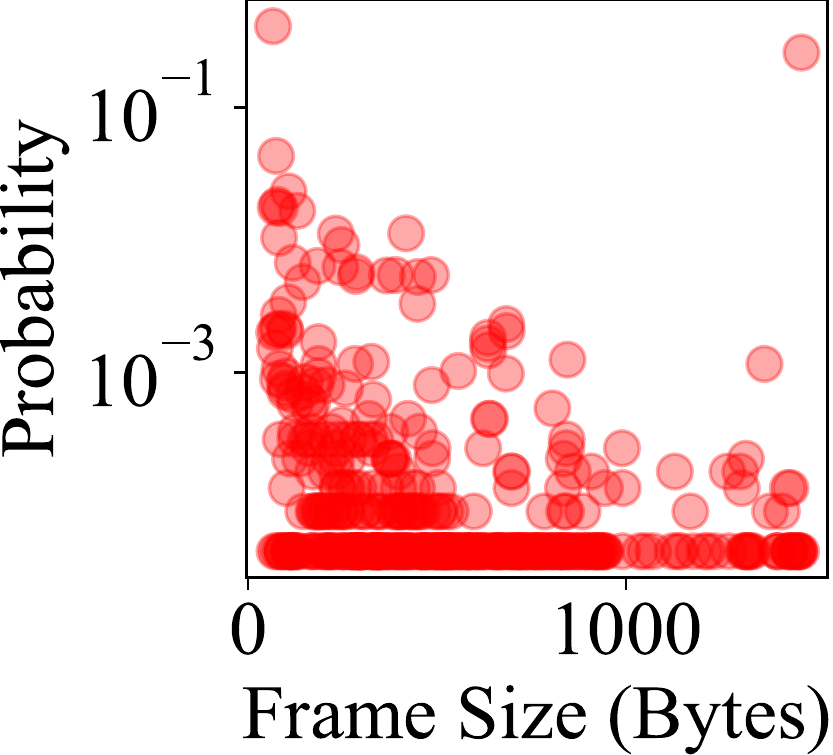}
		\label{figure_testcase_net_M1}
	} \quad
	\subfloat[Net\_M2]{
		\includegraphics[width = 0.13\linewidth]{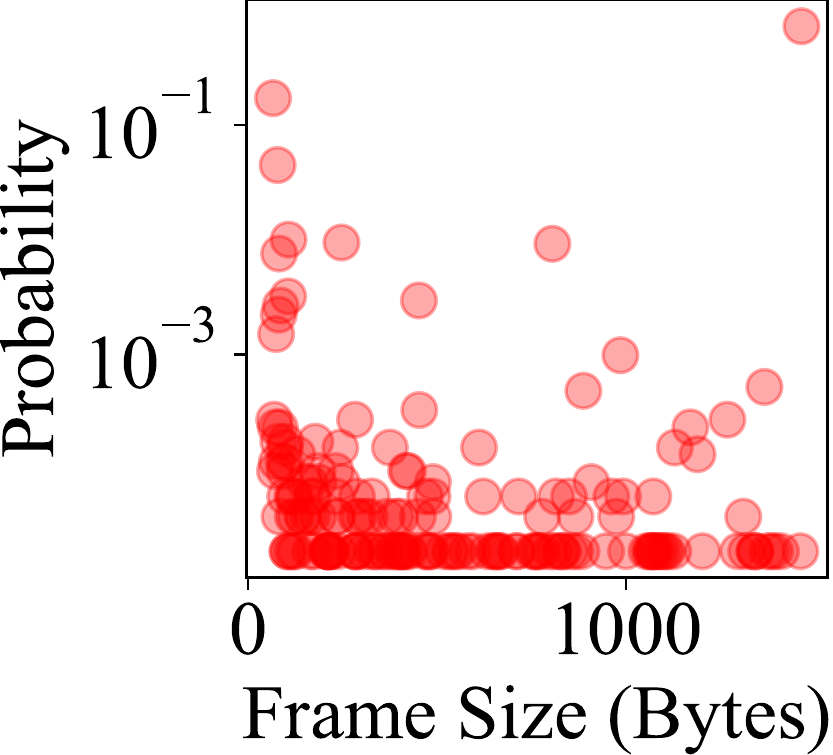}
		\label{figure_testcase_net_M2}
	} \quad        
	\subfloat[Net\_M3] {
		\includegraphics[width = 0.13\linewidth]{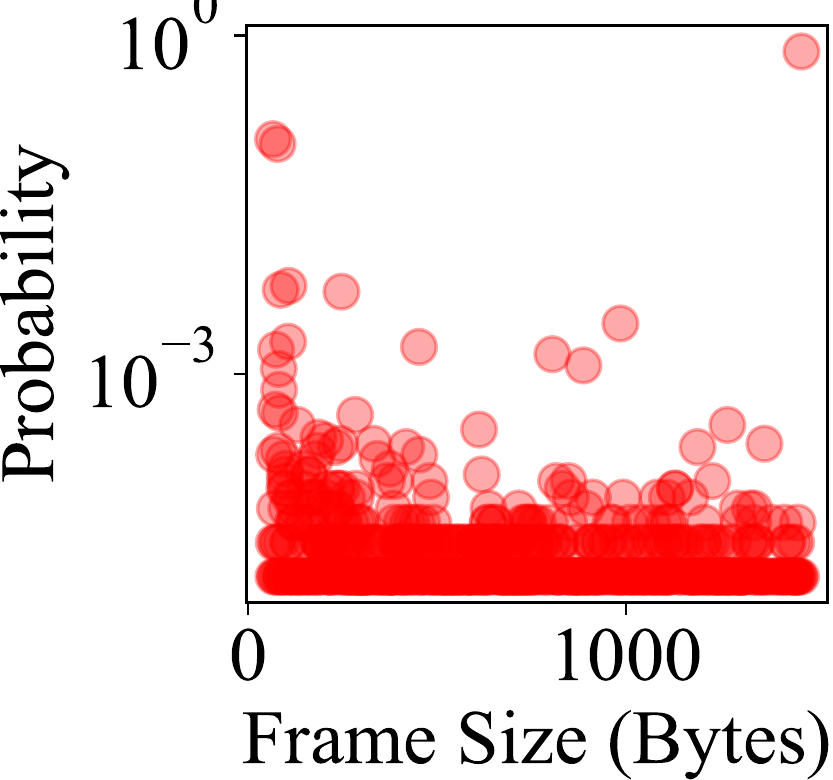}
		\label{figure_testcase_net_M3}
	} \quad
	\subfloat[Net\_M4] {
		\includegraphics[width = 0.13\linewidth]{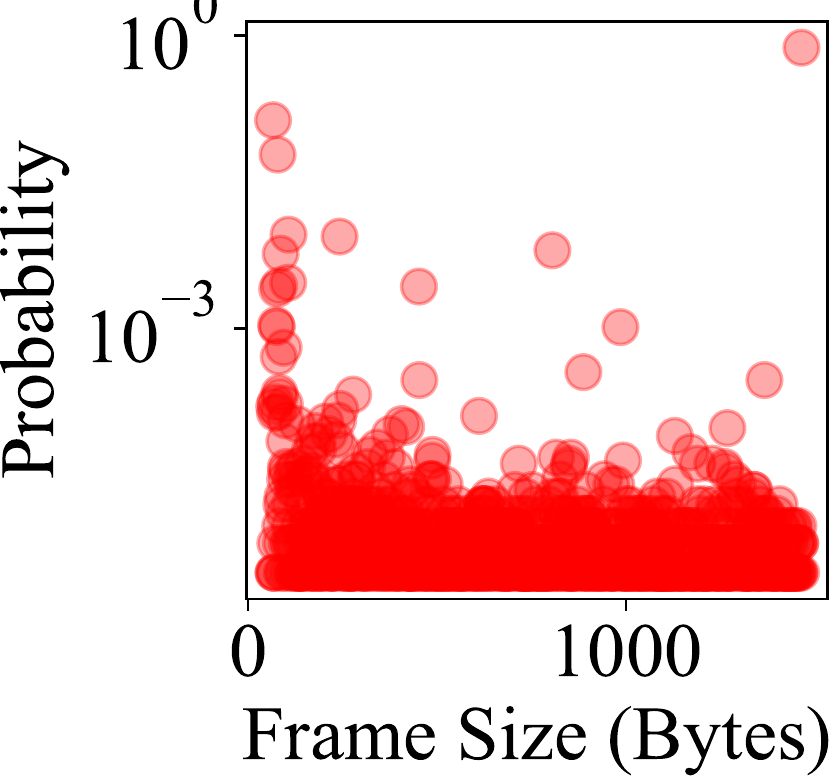}
		\label{figure_testcase_net_M4}
	} \quad
	\subfloat[Net\_M5]{
		\includegraphics[width = 0.13\linewidth]{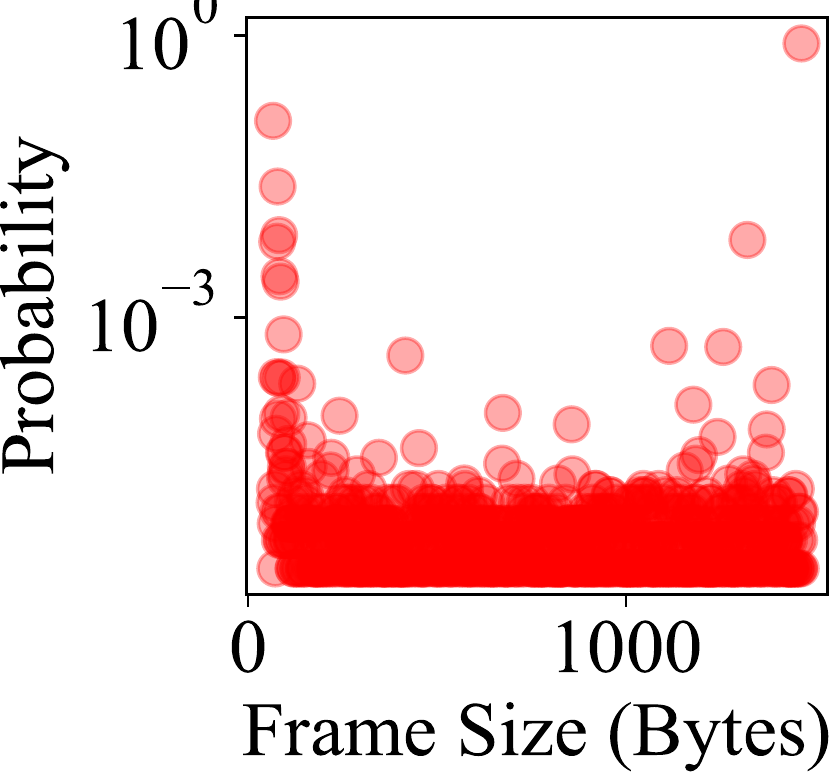}
		\label{figure_testcase_net_M5}
	} \quad
	\subfloat[Net\_M6]{
		\includegraphics[width = 0.13\linewidth]{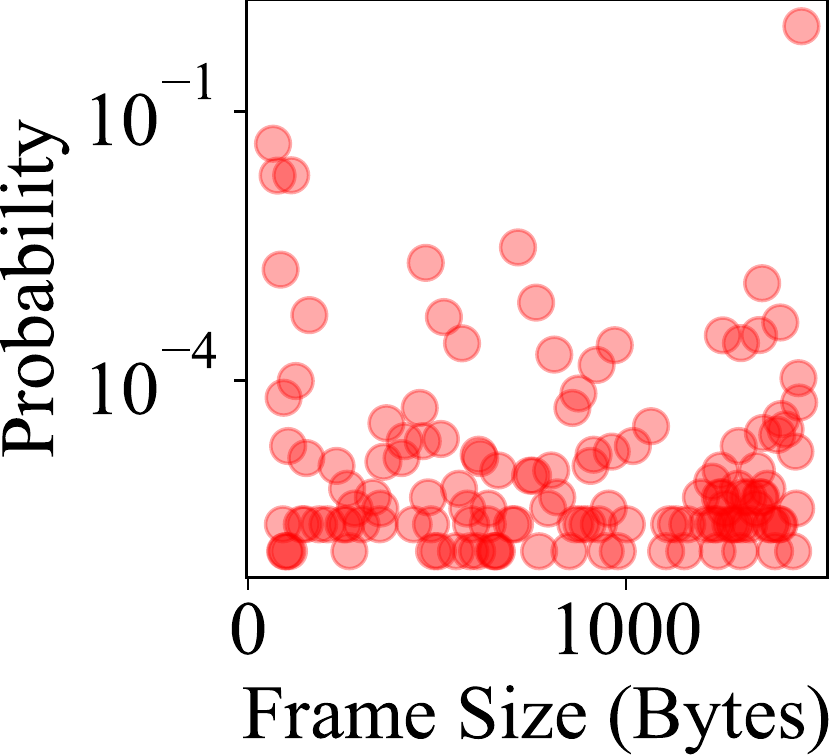}
		\label{figure_testcase_net_M6}
	} \\
	
	\subfloat[Net\_M7] {
		\includegraphics[width = 0.13\linewidth]{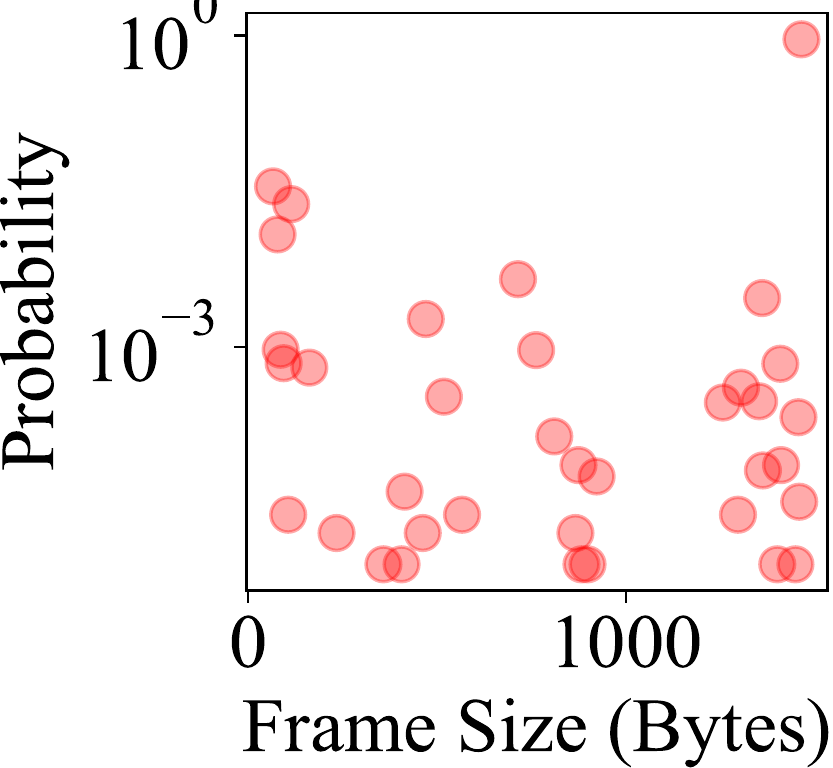}
		\label{figure_testcase_net_M7}
	} \quad
	\subfloat[Net\_M8] {
		\includegraphics[width = 0.13\linewidth]{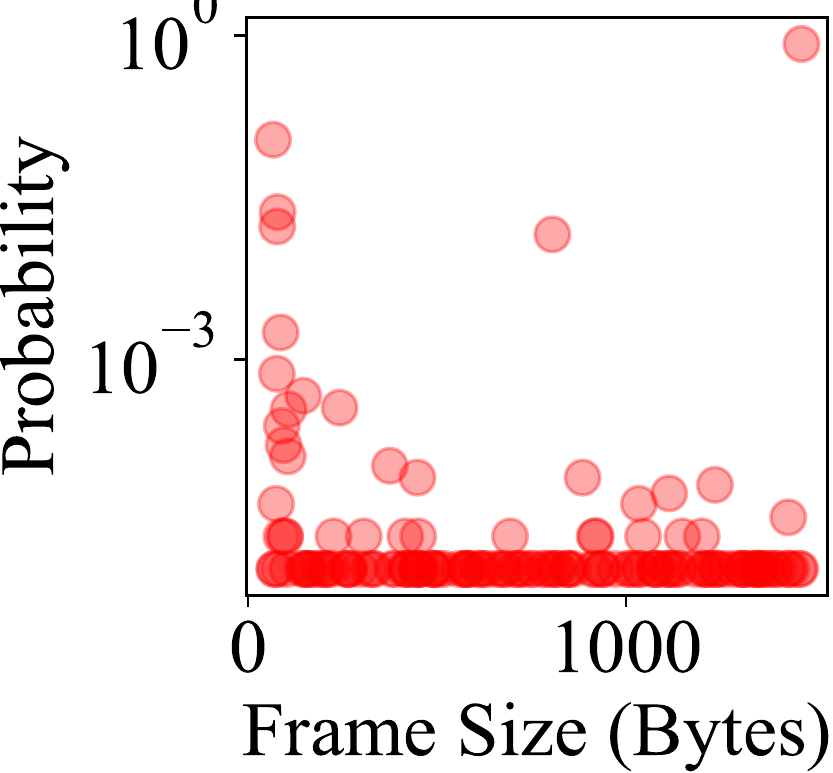}
		\label{figure_testcase_net_M8}
	} \quad 
	\subfloat[Net\_M9]{
		\includegraphics[width = 0.13\linewidth]{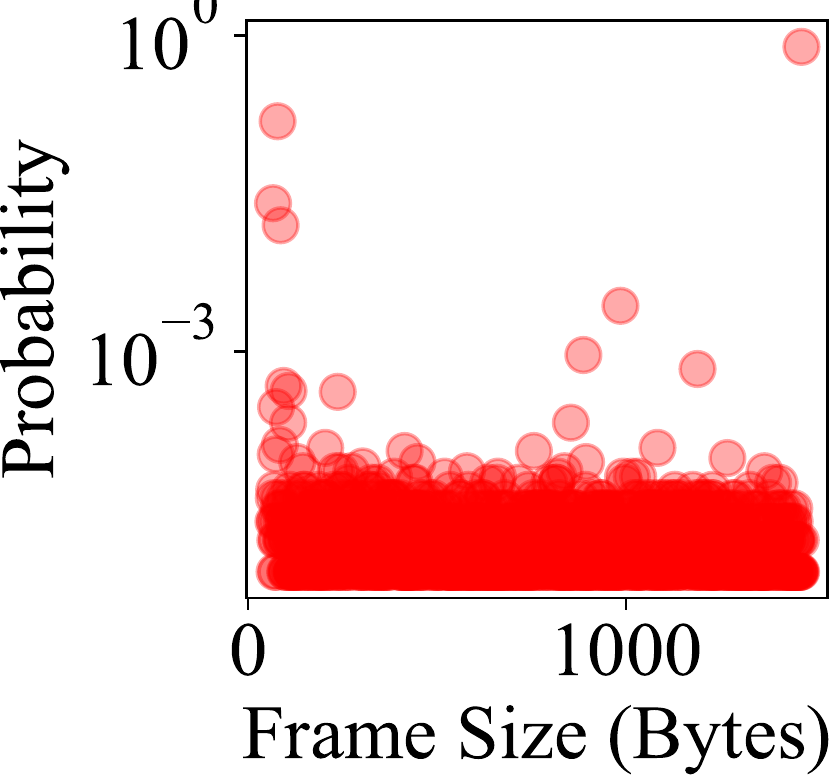}
		\label{figure_testcase_net_M9}
	} \quad
	\subfloat[Net\_M10]{
		\includegraphics[width = 0.13\linewidth]{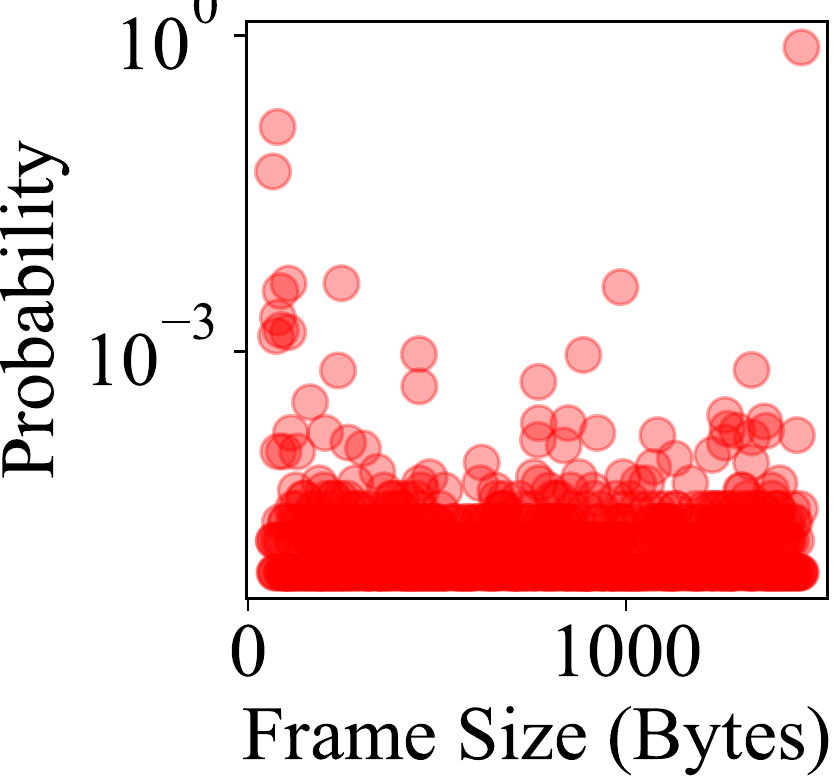}
		\label{figure_testcase_net_10}
	} \quad        
	\subfloat[Net\_M11] {
		\includegraphics[width = 0.13\linewidth]{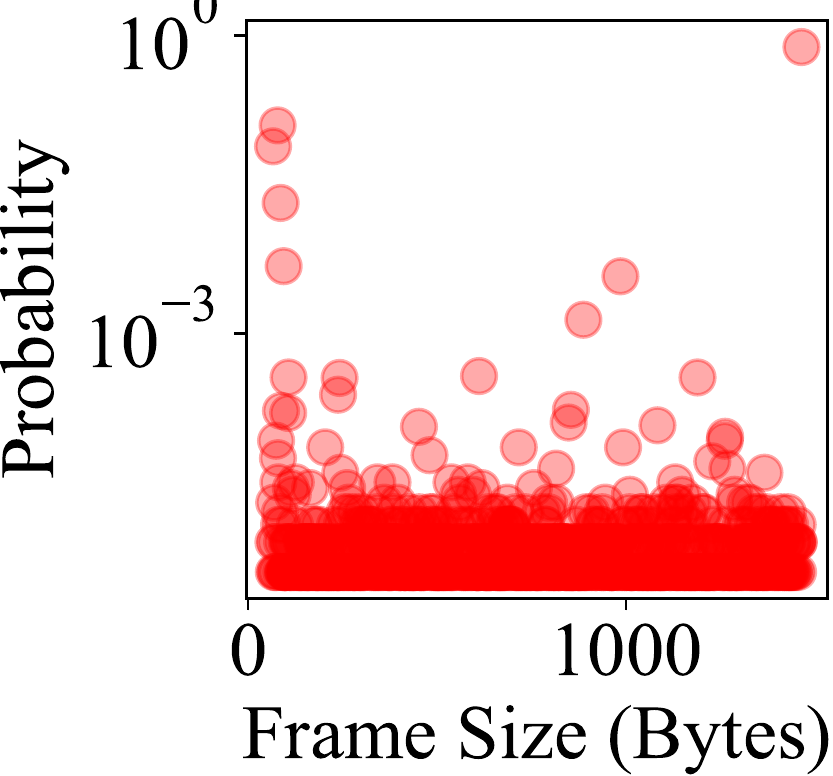}
		\label{figure_testcase_net_M11}
	} \quad
	\subfloat[Net\_M12] {
		\includegraphics[width = 0.13\linewidth]{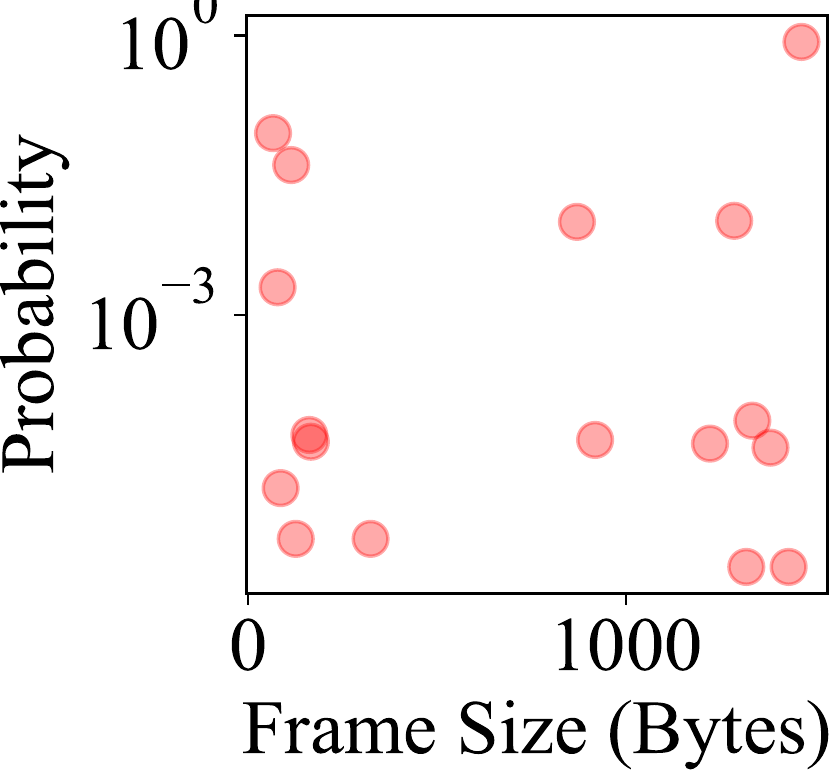}
		\label{figure_testcase_net_M12}
	} \\
	
	\caption{Packet Size Distribution of Micro Traces.}
	\label{figure_testcase_packet_all}
\end{figure*}

We further analyze the spatial patterns of the network traffic using a traffic matrix. Fig. ~\ref{figure_testcase_traffic_matrix_all} shows the traffic matrix patterns of all micro traces. Note that the darker the color, the more data transmission between the source-destination node pair. We find that they reflect diverse spatial patterns in terms of source-destination node pair and packet size.  

Packet size distribution is an important metric for data center networking. We analyze the packet size distribution of all micro traces, as shown in Fig. ~\ref{figure_testcase_packet_all}. We find that they have different probabilities for different scales of frame size. 

Above all, the micro traces represent a wide spectrum of network traffic patterns from the perspectives of time-series, spatial, and packet size patterns. Hence, DCNetBench has the ability to be scaled to different users and underlying technologies with different applications and configurations to fulfill different benchmarking requirements.

\subsection{Switch Chip  Evaluation}~\label{switch-eva}

We further use micro traces to evaluate five switch chips comprehensively from different perspectives: Average Forwarding Performance (AF), Worst Forwarding Performance (WF), Latency Jitter (LJ), Congestion Control Performance (CC), and Burst Absorption (BA). 
For each perspective, we select the corresponding micro traces and the representative metric, as shown in Table \ref{table_benchmark_indices}.

\textbf{Average Forwarding Performance (AF)}: We use the average latency of all packets as the metric to measure the average forwarding capability of the device. We replay the micro traces from Net\_M9 to Net\_M12 with stable patterns.

\textbf{Worst Forwarding Performance (WF)}: We use the 99th percentile latency of all packets to reflect the worst forwarding capability. We replay the micro traces from Net\_M5 to Net\_M8 with increase patterns.

\textbf{Latency Jitter (LJ)}: We use the standard deviation of the latency to measure the stability of the forwarding capability. We replay the micro traces from Net\_M5 to Net\_M8 with increase patterns.

\textbf{Congestion Control Performance (CC)}: We use the packet loss rate during the whole execution time to reflect the congestion control performance. We replay the micro traces from Net\_M1 to Net\_M4 with burst patterns.

\textbf{Burst Absorption (BA)}: Different from CC, we compute the packet loss rate every 50 milliseconds and use the standard deviation of these packet loss rates to measure the burst absorption performance. We replay the micro traces from Net\_M1 to Net\_M4 with burst patterns.





\begin{table}[htb]
	\caption{Metrics and Traces for Switch Chip Evaluation. AVG refers to the average and STD refers to the standard deviation.}
	\label{table_benchmark_indices}
	\resizebox{1\columnwidth}{!} {
		\begin{tabular}{ccc}
			\toprule
			\textbf{Perspective}               &  \textbf{Metric}                         & \textbf{Traces}         \\
			\midrule
			Average Forwarding Performance (AF)          & AVG(Latency)                  &  Net\_M9, Net\_M10, Net\_M11, Net\_M12  \\
			Worst Forwarding Performance (WF)        & 99th Percentile Latency            &  Net\_M5, Net\_M6, Net\_M7, Net\_M8  \\
			Latency Jitter (LJ)                  & STD(Latency)                  &  Net\_M5, Net\_M6, Net\_M7, Net\_M8   \\
			Congestion Control Performance (CC)              & Packets Loss Rate             &  Net\_M1, Net\_M2, Net\_M3, Net\_M4   \\
			Burst Absorption (BA)         & STD(Packets Loss Rate)   &  Net\_M1, Net\_M2, Net\_M3, Net\_M4  \\
			
			\bottomrule                         
		\end{tabular}
	}
\end{table}

\begin{figure}[htb]
			\centering
			\includegraphics[width=0.45\textwidth]{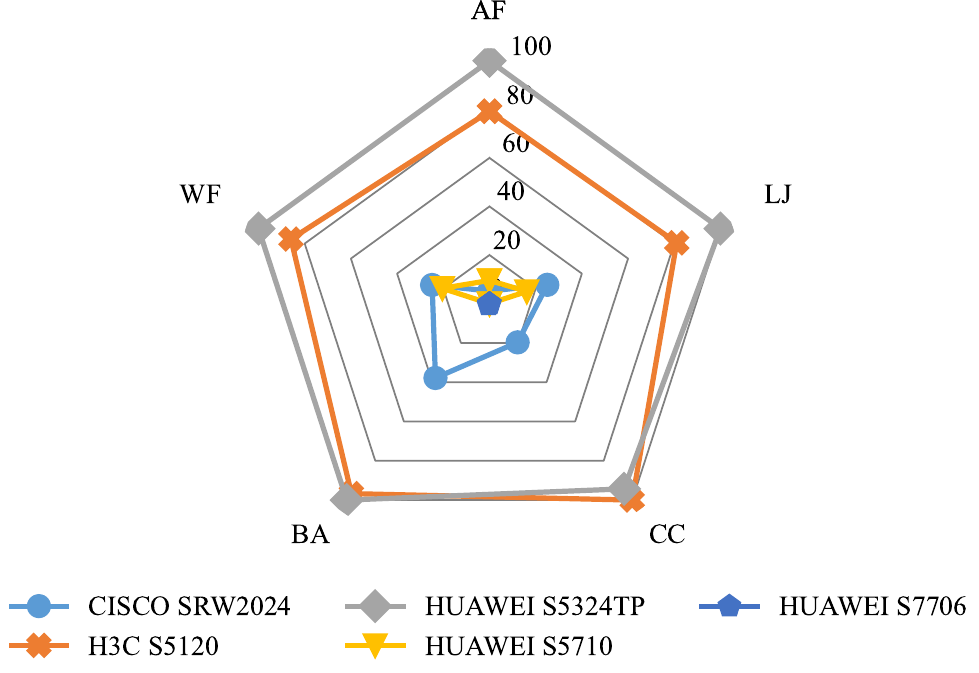}
			\caption{The Comparison of Five Switches. The smaller the number, the better the performance.}
			\label{figure_benchmarking_result_radar}
		\end{figure}

		The evaluation results are shown in 
		Fig.~\ref{figure_benchmarking_result_radar}.
		We find that HUAWEI S7706 has the optimal or near-optimal performance from all perspectives, which also corresponds to the rated parameters of the device in Table ~\ref{table_evaluation_device_info} (Its exchange capacity and sending speed are one to two orders of magnitude higher than other switches). HUAWEI S5710, with the second-best-rated parameters,  also achieves decent performance. From Table ~\ref{table_evaluation_device_info}, CISCO SRW2024 and HUAWEI S5324TP have the worst exchange capacity and sending speed. However, according to our evaluation, CISCO SRW2024 achieves comparable performance to HUAWEI S5710 in the perspectives of Average Forwarding Performance, Latency Jitter, and Worst Forwarding Performance. It outperforms H3C S5120 and HUAWEI S5324TP from all performance perspectives. The result implies: (1) the rated parameters are not enough to reflect the comprehensive capacity of the switches, and (2) we should choose the appropriate device, rather than the best device, according to the traffic patterns of the applications. 
		
		
		
		
\section{Conclusion}

This article proposed that it is essential to scale different DC workloads and underlying technologies in designing, implementing, and evaluating DC networking benchmarking. We built DCNetBench, the first application-driven data center network benchmarking that can scale to different users, underlying technologies, and varying benchmarking requirements. We built a MaxiNet-based emulated system that can simulate networking with different configurations. Then we ran applications on the emulated systems to capture the realistic network traffic patterns; we analyzed and classified these patterns to model and replay those traces. Finally, We provided an automatic benchmarking framework to support this pipeline. The evaluations on DCNetBench show its scaleability, effectiveness, and diversity for data center network benchmarking.


\section*{Acknowledgment}
This work does not raise any ethical issues.

\bibliographystyle{ieeetr}

\end{document}